\documentclass[a4paper,aps,pre,floatfix,twocolumn,showpacs]{revtex4}

\usepackage{epsfig}
\usepackage{graphicx}
\usepackage{latexsym,amsmath,verbatim}

\newcommand{\avk}{\langle k\rangle}

\begin{document}

\title{Random walks on complex trees}

\author{Andrea Baronchelli, Michele Catanzaro, and Romualdo
  Pastor-Satorras} 
\affiliation{Departament de F\'\i sica i Enginyeria Nuclear,
  Universitat Polit\`ecnica de Catalunya, Campus Nord B4, 08034
  Barcelona, Spain}

\date{\today}

\begin{abstract}
  We study the properties of random walks on complex trees. We observe
  that the absence of loops reflects in physical observables showing
  large differences with respect to their looped counterparts. First,
  both the vertex discovery rate and the mean topological displacement
  from the origin present a considerable slowing down in the tree
  case.  Second, the mean first passage time (MFPT) displays a
  logarithmic degree dependence, in contrast to the inverse degree
  shape exhibited in looped networks. This deviation can be ascribed
  to the dominance of source-target topological distance in trees. To
  show this, we study the distance dependence of a symmetrized MFPT
  and derive its logarithmic profile, obtaining good agreement with
  simulation results. These unique properties shed light on the
  recently reported anomalies observed in diffusive dynamical systems
  on trees.
\end{abstract}

\pacs{89.75.Hc, 05.40.Fb, 05.60.Cd}

\maketitle

\section{Introduction}
\label{sec:introduction}

Diffusion problems on tree structures pop up in a wide range of
scientific domains, such as theoretical
physics~\cite{huberman1985,huberman1986,sibani1989}, computer
science~\cite{lee_microcomputer,card_book}, phylogenetic
analysis~\cite{cavallisforza1967} and cognitive
science~\cite{fisher_machine_learning}.  Moreover, dynamics in tree
structures have gained a renewed interest in the physics community as
a spin-off of the attention devoted to the structural properties of
complex networks~\cite{barabasi02,mendesbook} and dynamical processes
taking place on top of them~\cite{dorogovtsev_critical2007}.  Thus,
along with the widely explored scale-free (SF)
networks~\cite{barabasi_albert_first}, also SF trees have started to
be used as underlying topologies for dynamical
processes. Interestingly, the absence of loops in trees turns out to
have a strong impact on the considered dynamics, and relevant
differences between looped networks and tree topologies have been
recently reported in several dynamical models, such as the voter
model~\cite{castellano_voter2}, the naming
game~\cite{dallasta_ng_nets}, the random walk and the
pair-annihilation processes~\cite{nohkim}, and a model for norm
spreading~\cite{nakamaru2004stl}.  The properties of most dynamical
processes on looped network can be reasonably accounted for by
annealed mean-field theories \cite{dorogovtsev_critical2007}, which
rely only on information about the degree distribution and degree
correlations \cite{serrano07:_correl}, and consider the network as
maximally random at all other respects. The behavior observed in
trees, different from the annealed mean-field predictions, must thus
be explained in terms of the non-local constrain of absence of loops
imposed in this kind of graphs, which is hard to implement in
theoretical approaches.

In this paper we explore the peculiarities induced in dynamical
processes by the absence of loops by considering the simplest possible
example, namely the uncorrelated random walk~\cite{hughes,lovasz}.
Several works have been devoted in the past to the study of random
walks on complex networks, showing in general a good agreement between
theory and simulations on looped networks, while differences were
reported in tree networks in Ref.~\cite{nohkim}. Here, we find that
the global constraint of lack of loops induces a general slowing down
of diffusion, as measured by the network coverage and the mean
topological displacement from the origin. As well, it profoundly
alters the degree dependence of the mean-first passage time. This is
due to the fact that the source-target distance is dominating in
trees. In order to account for this features, we study the mean round
trip time versus distance and find an analytic expression of its
dependence on degree.

\section{Random walks on complex networks and trees}
\label{sec:random-walks-random}

We consider random walks on general networks defined by a walker that,
located on a given vertex of degree $k$ at time $t$, hops with
probability $1/k$ to one of the $k$ neighbors of that vertex at time
$t+1$. We have measured the properties of random walks on growing SF
trees created with the linear preferential attachment (LPA)
algorithm~\cite{barabasi_albert_first,mendes99}: at each time step
$s$, a new vertex with $m$ edges is added to the network and connected
to an existing vertex $s'$ of degree $k_{s'}$ with probability $\Pi_{s
  \rightarrow s'} = (k_{s'}+a)/(2m+a)s$. This process is iterated
until reaching the desired size $N$. The resulting network has degree
distribution $P(k) \sim k^{-\gamma}$ with tunable exponent $\gamma=3 +
a/m$, with $\gamma<3$ for $a<0$. For $m=1$ the LPA model yields a
strict tree topology.  Degree correlations, measured by the average
degree of the nearest neighbors of the vertices of degree $k$
\cite{alexei}, are given by $\bar{k}_{nn}(k) \sim
N^{(3-\gamma)/(\gamma-1)}
k^{-3+\gamma}$~\cite{barrat2005rea}. Therefore, only for $\gamma=3$
($a=0$) we expect to obtain uncorrelated networks.

In order to explore the intrinsic properties of a tree topology,
disregarding SF effects, we have also considered homogeneous networks.
In the growing exponential network model (EM) \cite{mendesbook}, at
each time step $s$ a new vertex with $m$ edges is added to the
network, and it is connected to $m$ randomly chosen other vertices. In
the continuous degree approximation (i.e., considering the degree as a
continuous variable and substituting sums by integrals), this models
leads to networks with an exponential degree distribution, $P(k) =
e^{1-k/m}/m$. Again, homogeneous trees are generated by selecting
$m=1$. The random Cayley tree (RC), on the other hand, is generated by
adding $z$ neighbors to a randomly selected leaf (i.e. a vertex whose
degree is $k=1$) at each time step $s$ ($z+1$ neighbors are added to
the first vertex). The resulting tree contains only vertices with
degree $k=z+1$, and leaves with $k=1$.
 
To check our results against looped structures, we have considered the
uncorrelated configuration model (UCM)~\cite{catanzaro_ucm}, yielding
uncorrelated networks with any prescribed SF degree distribution. The
model is defined as follows: (1) Assign to each vertex $i$ in a set of
$N$ initially disconnected vertices a degree $k_i$, extracted from the
probability distribution $P(k) \sim k^{-\gamma}$, and subject to the
constraints $m \leq k_i \leq N^{1/2}$ and $\sum_i k_i$ even. (2)
Construct the network by randomly connecting the vertices with $\sum_i
k_i /2$ edges, respecting the preassigned degrees and avoiding
multiple and self-connections. Using this algorithm, it is possible to
create SF networks which are completely uncorrelated. Additionally, by
selecting the minimum degree $m \geq 2$, we generate connected
networks with probability almost $1$.  The effect of correlations in
looped structures can be checked by means of the configuration model
(CM)~\cite{mendesbook}, which is analogous to the UCM, but allows
degrees to range in the interval $m \leq k_i \leq N$
\cite{mariancutofss}.  In all present simulations, we set for looped
neworks $m=4$, tree networks corresponding to $m=1$ ($z=4$ for the RC
tree).

\section{Random walk exploration}
\label{sec:rand-walk-expl}

We start by studying two properties of a random walk that quantify the
speed at which it explores its neighborhood in the network. The first
one is the coverage $S(t)$, defined as the number of different
vertices visited by a walker at time $t$, averaged for different
random walks starting from different sources. For looped networks, the
coverage reaches after a short transient the functional form
\cite{stauffer_annealed2005,almaas03:_scaling} $S_L(t) \sim t$
\footnote{In the following, the subscripts $L$ and $T$ will indicate
  looped and tree networks, respectively.}, in accordance with
theoretical calculations for the Bethe lattice \cite{hughes82}, and
eventually saturates to $S_L(\infty) = N$, due to finite size
effects. A scaling form for the coverage has been proposed
\cite{stauffer_annealed2005} to be $S_L(t) = N f(t/N)$, with $f(x)
\sim x$ for $x\ll 1$ and $f(x) \sim 1$ for $x\gg 1$.

\begin{figure}[!t]
  \centerline{
    \includegraphics*[width=0.40\textwidth]{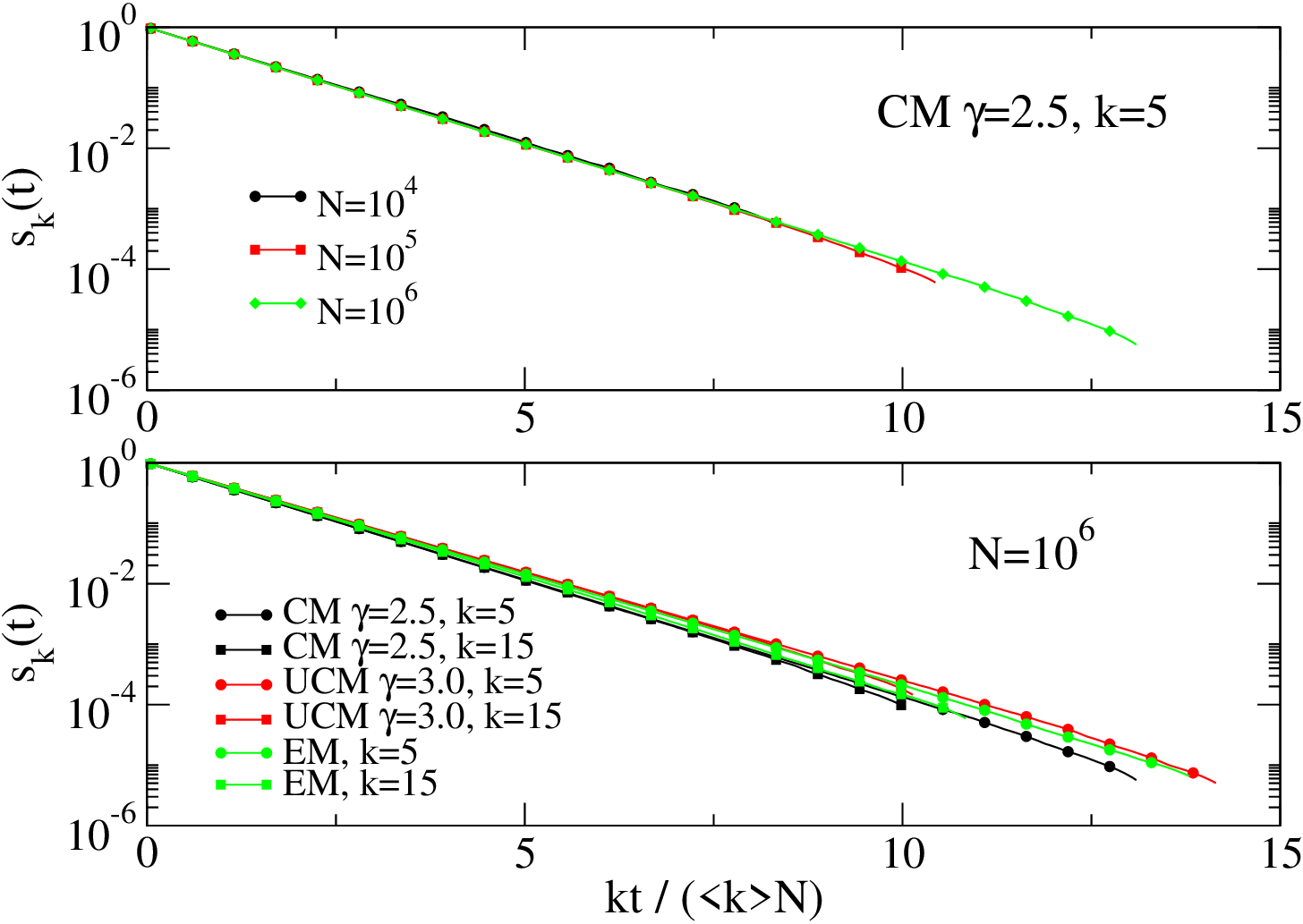}
  }	
  \caption{(Color online) Coverage spectrum $s_k(t)$ in looped complex
    networks as a function of $k t / \avk N$. Top: Curves for the same
    degree and different network size $N$.  Bottom: Curves for
    different degrees and fixed network size.}
\label{f:coverageL}
\end{figure}

The origin of the scaling of the coverage with system size can be
understood by means of a simple dynamic mean-field argument. Let us
define $\rho_k(t)$ as the probability that a vertex of degree $k$
hosts the random walker at time $t$.  During the evolution of the
random walk, this probability satisfies, in a general network with a
correlation pattern given by the conditional probability $P(k'|k)$
that a vertex of degree $k$ is connected to another vertex of degree
$k'$ \cite{serrano07:_correl}, the mean-field equation
\begin{equation}
  \frac{\partial \rho_k(t)}{\partial t} = -\rho_k(t) + k \sum_{k'}
  \frac{P(k'|k)}{k'} \rho_{k'}(t).
\end{equation}
In the steady state, $\partial_t \rho_k(t)=0$, the solution of this
equation, for any correlation pattern, is given by the normalized
distribution \cite{lovasz,nohrieger}
\begin{equation}
  \rho_k(t) = \frac{k}{\avk N}.
  \label{eq:2}
\end{equation}
Let us now define the coverage spectrum $s_k(t)$ as fraction of
vertices of degree $k$ visited by the random walker at least
once. Obviously, we have that $S(t) = N \sum_k P(k) s_k(t)$. The
spectrum $s_k(t)$ increases in time as the random walk arrives to
vertices that have never been visited. Therefore, at a mean-field
level, it fulfills the rate equation
\begin{equation}
  \frac{\partial s_k(t)}{\partial t} = k [1-s_k(t)] \sum_{k'}
  \frac{P(k'|k)}{k'} \rho_{k'}(t).
\end{equation}
Approximating $\rho_{k'}(t)$ by its steady-state value (for not too
small times), we obtain
\begin{equation}
  \frac{\partial s_k(t)}{\partial t} = [1-s_k(t)] \frac{k}{\avk N},
\end{equation}
whose solution, with the initial condition $s_k(0)=0$ is
\begin{equation}
  s_k(t) = 1 - \exp\left(-  \frac{k t}{\avk N} \right).
  \label{eq:1}
\end{equation}
We therefore are lead to the general scaling expression
\begin{equation}
  \frac{S(t)}{N} = 1 - \sum_k P(k)  \exp\left(-  \frac{k t}{\avk N}
  \right). 
\end{equation}
In the limit of $k t / \avk N \ll 1$, we recover the exact result
$S(t) \sim t$ \cite{hughes82}.  For SF networks, we obtain within the
continuous degree approximation
\begin{eqnarray}
  \frac{S(t)}{N} &=& 1 - (\gamma-1)m^{\gamma-1} \int_m^\infty
  k^{-\gamma}  \exp\left(-  \frac{k t}{\avk N} 
  \right)\; dk \nonumber \\
  &=& 1-(\gamma-1) E_\gamma \left(\frac{ m t }{\avk N} 
  \right),\label{eq:3}
\end{eqnarray}
where $E_\gamma(z)$ is the exponential integral function
\cite{abramovitz}. For EM networks, on the other hand, we find
\begin{eqnarray}
  \frac{S(t)}{N} &=& 1 - \frac{e}{m} \int_m^\infty
  e^{-k/m}  \exp\left(-  \frac{k t}{\avk N} 
  \right)\; dk \nonumber \\
  &=& 1-\frac{e^{-m t / \avk N}}{1 + \frac{m t}{\avk N}}. \label{eq:4} 
\end{eqnarray}

\begin{figure}[!t]
  \centerline{
    \includegraphics*[width=0.40\textwidth]{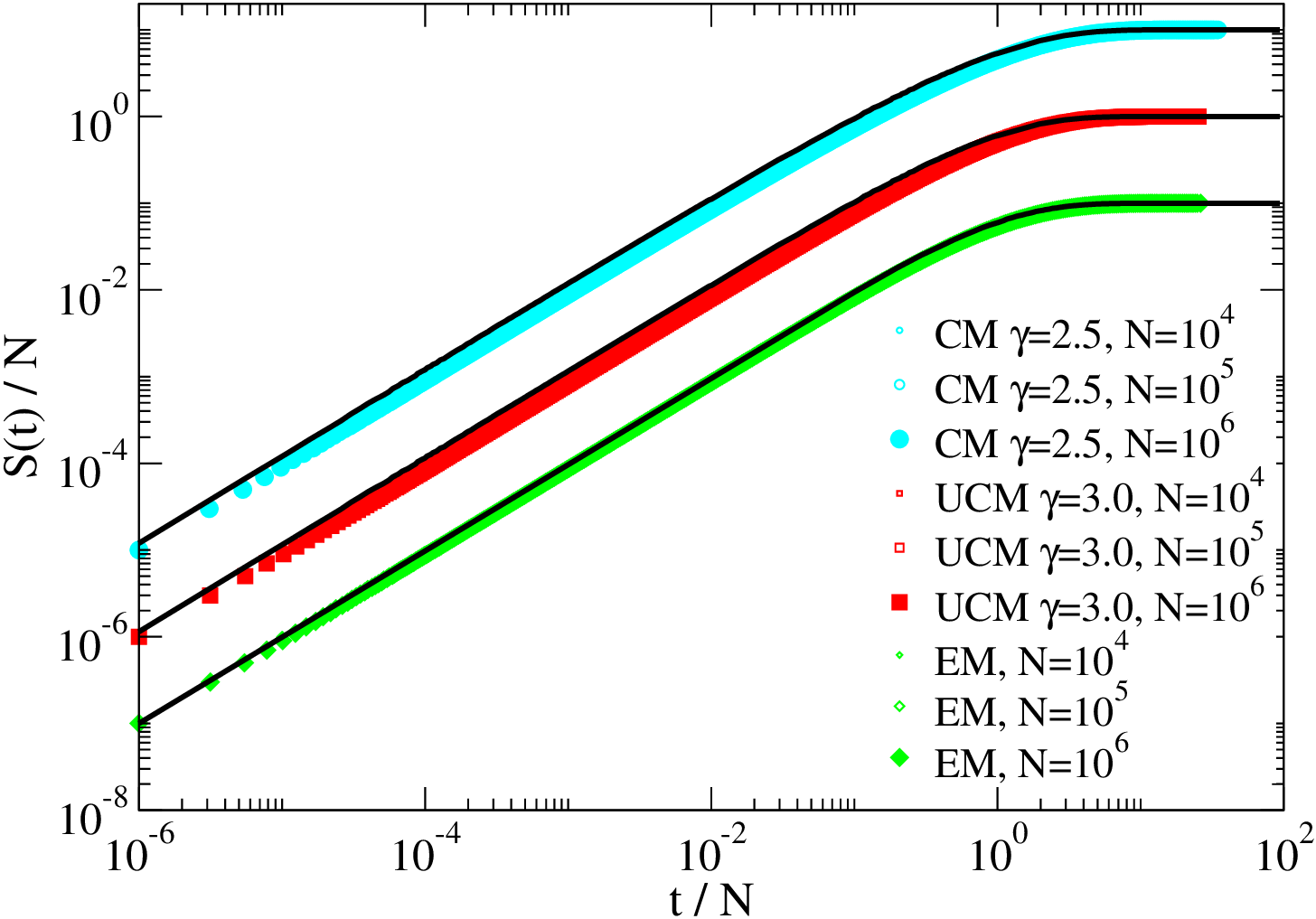}
  }
  \caption{(Color online) Rescaled coverage for looped complex
    networks.  We plot in full lines the analytical predictions
    Eqs.~(\ref{eq:3}) and~(\ref{eq:4}), corresponding to SF and EM
    networks. Results for CM, UCM, and EM networks have been shifted
    (top to bottom) in the vertical axis for clarity.}
  \label{f:coverageLtotal}
\end{figure}

In Fig.~\ref{f:coverageL}, we can observe that the scaling predicted
by Eq.~(\ref{eq:1}) for the coverage spectrum $s_k(t)$ is very well
satisfied in looped complex networks, independently of their
homogeneous or SF nature, and in this last case, of the degree
exponent and the presence or absence of correlations. In
Fig.~\ref{f:coverageLtotal}, on the other hand, we plot the total
coverage $S(t)/N$, which can be fitted quite correctly by the
analytical expressions Eqs.~(\ref{eq:3}) and~(\ref{eq:4}) for SF and
EM networks.

\begin{figure}[!t]
  \centerline{
    \includegraphics*[width=0.40\textwidth]{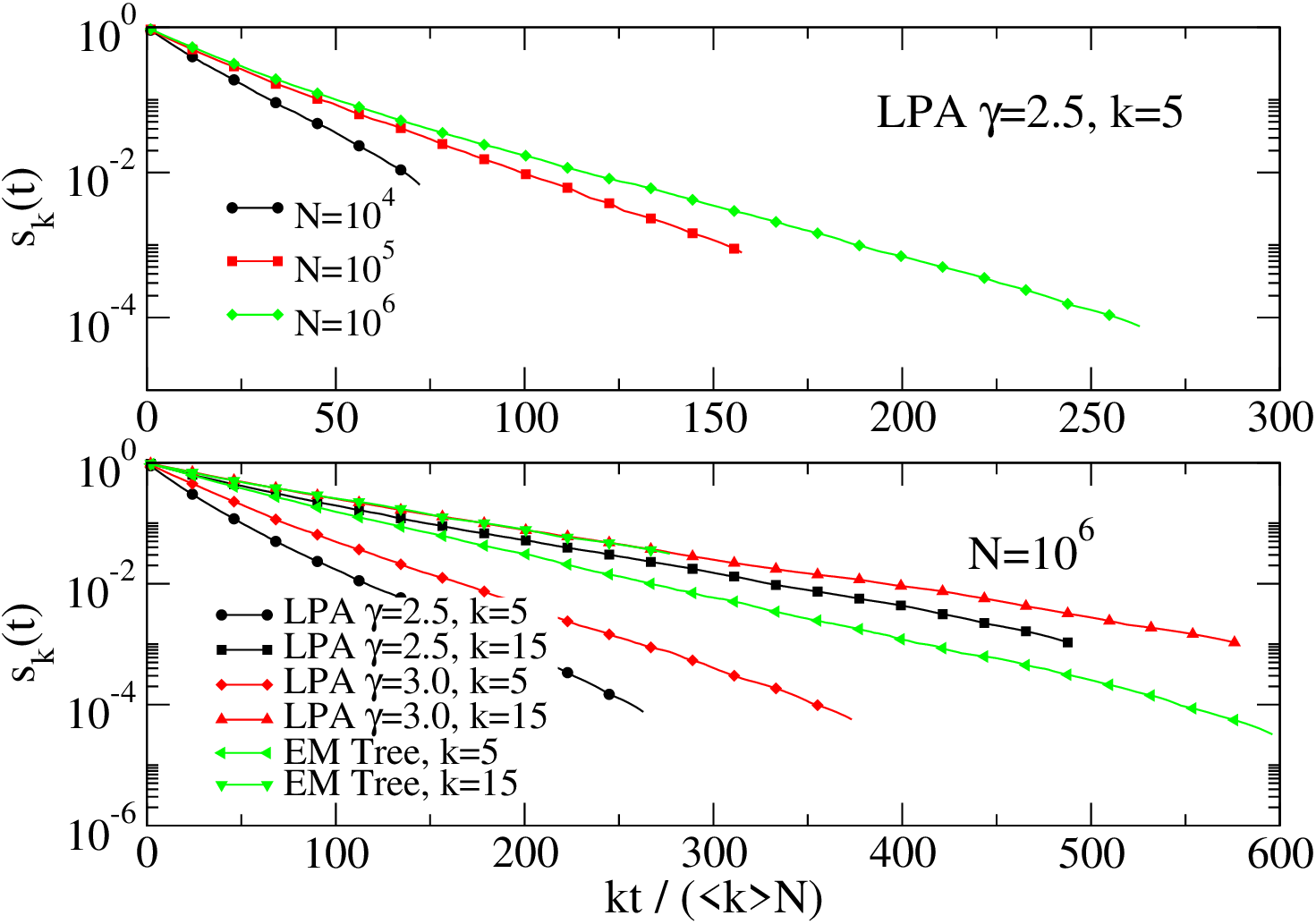}
  }	
  \caption{(Color online) Coverage spectrum $s_k(t)$ in tree networks
    as a function of $k t / \avk N$.  Top: Curves for the same degree
    and different network size $N$.  Bottom: Curves for different
    degrees and fixed network size. The scaling here is different from
    the mean-field prediction Eq.~(\ref{eq:1}).}
  \label{f:coverageT}
\end{figure}

\begin{figure}[!t]
  \centerline{
    \includegraphics*[width=0.40\textwidth]{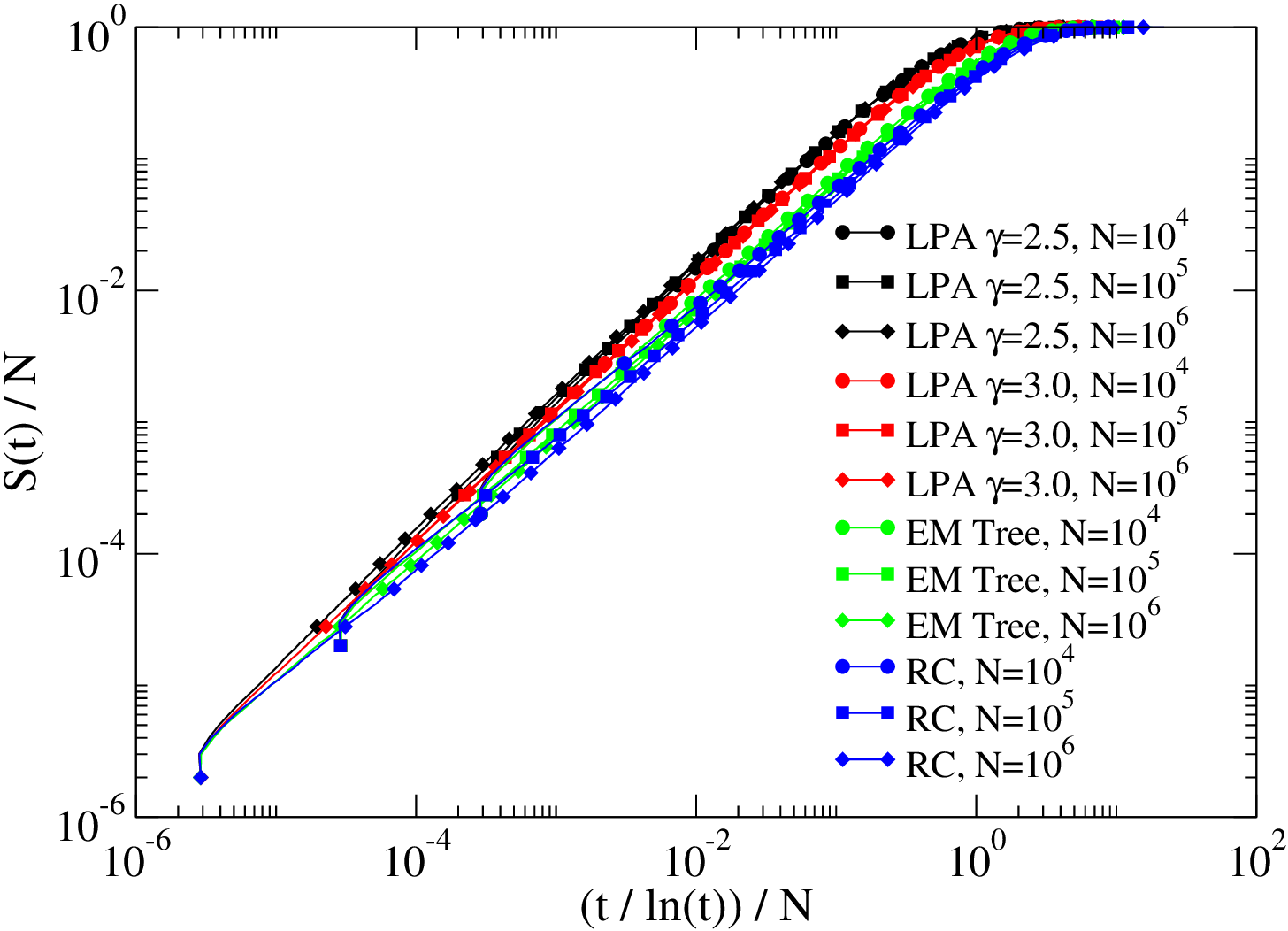}
  }
  \caption{(Color online) Rescaled coverage as a function of time in
    complex trees. Results for LPA ($\gamma=2.5$), LPA ($\gamma=3.0$),
    EM, and RC networks have been shifted (top to bottom) in the
    vertical axis for clarity.}
\label{f:coverageTtotal}
\end{figure}

\begin{figure}[!t]
  \centerline{
    \includegraphics*[width=0.40\textwidth]{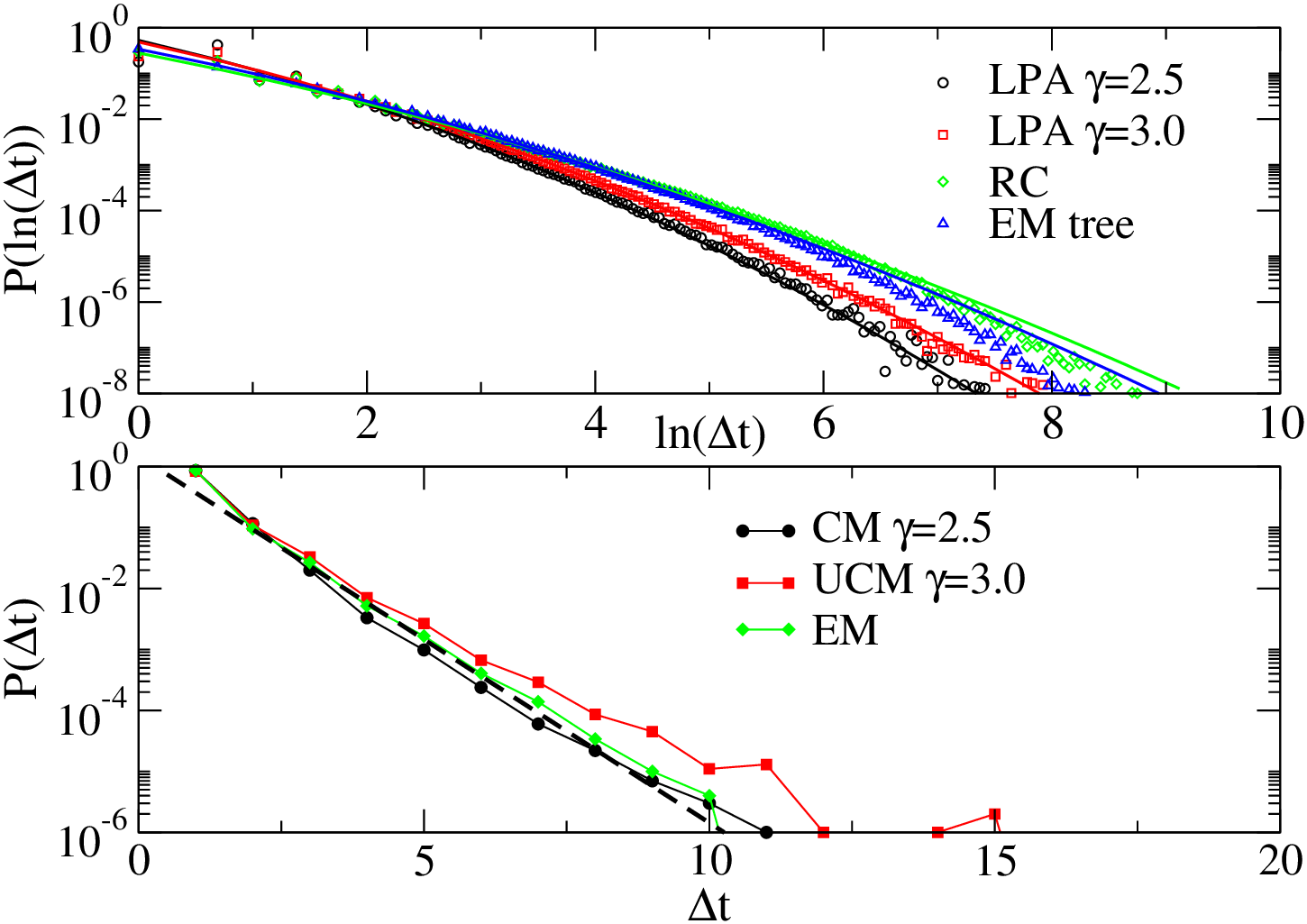}
  }	
  \caption{(Color online) Distribution of lag times in tree (top) and
    looped (bottom) networks can be fitted, respectively, by a
    log-normal (full lines) and an exponential (dashed line)
    distribution.  Data refer to graphs of size $N=10^5$.}
  \label{f:timeLags}
\end{figure}

On tree networks we find a different scenario.  In
Fig.~\ref{f:coverageT} we can see that the coverage spectrum does not
scale as predicted by our mean-field argument.  While we do not have
theoretical predictions for the correct scaling form, a numerical
analysis of the total coverage, Fig.~\ref{f:coverageTtotal}, shows
that, at short times, it grows in trees as $S_T(t) \sim t/ \ln(t)$,
preserving an approximate scaling form
\begin{equation}
  S_T(t) = N f\left(\frac{t}{\ln(t)N} \right),
\end{equation}
with a scaling function $f(x)$ that depends slightly on the network
details (degree exponent, correlations, etc.).  This observation
indicates the presence of a general slowing down mechanism in the
random walk dynamics in trees: the dynamics turns out to be more
recurrent and therefore it is more costly to find new vertices during
the walk. It is easy to see that this situation will correspond to a
walker deep in the leaves of a subtree that has otherwise completely
explored. In order to find new vertices, the walker must first find
the exit to the subtree.
This difficulty in finding new vertices can be directly measured by
the time lag $\Delta t$ between the discovery of two new vertices. In
Fig.~\ref{f:timeLags} we plot the probability distribution of time
lags, $P(\Delta t)$, computed for the discovery of the first $1\%$ of
the network, for looped and tree structures. We observe that, in
looped networks, this distribution takes an exponential form,
compatible with an almost constant time lag between the discovery of
two new vertices. In tree networks, this distribution shows instead
long tails, that can be fitted to a lognormal form, indicating that,
in some events (i.e. when the walker is trapped in one leaf in a
subtree) the discovery of a new vertex can take an unusually large
time.

Acute signatures of slowing down can be found also in the analysis of
the mean topological displacement (MTD) $\bar{d}(t)$ of the walker
from its origin at time $t$, defined as the shortest path length from
the vertex of origin of the random walk, to the vertex it occupies at
time $t$, averaged over different source vertices, and different
random walk realizations. In other works, the mean square topological
displacement (RMSTD) $\bar{d^2}(t)$
\cite{almaas03:_scaling,gallos04:_random_walk} was instead
considered. In complex networks, and since the shortest path length is
a positive definite quantity, both quantities yield the same scaling
result, i.e.  $\bar{d^2}(t) \sim [\bar{d}(t)]^2$ \footnote{Differences
  can however appear in networks with and underlying metric space,
  such as the Watts-Strogatz network \cite{watts_strogatz} for
  rewiring parameter $p \ll 1$, see Ref.~\cite{almaas03:_scaling}.}.

\begin{figure}[t]
  \centerline{
    \includegraphics*[width=0.40\textwidth]{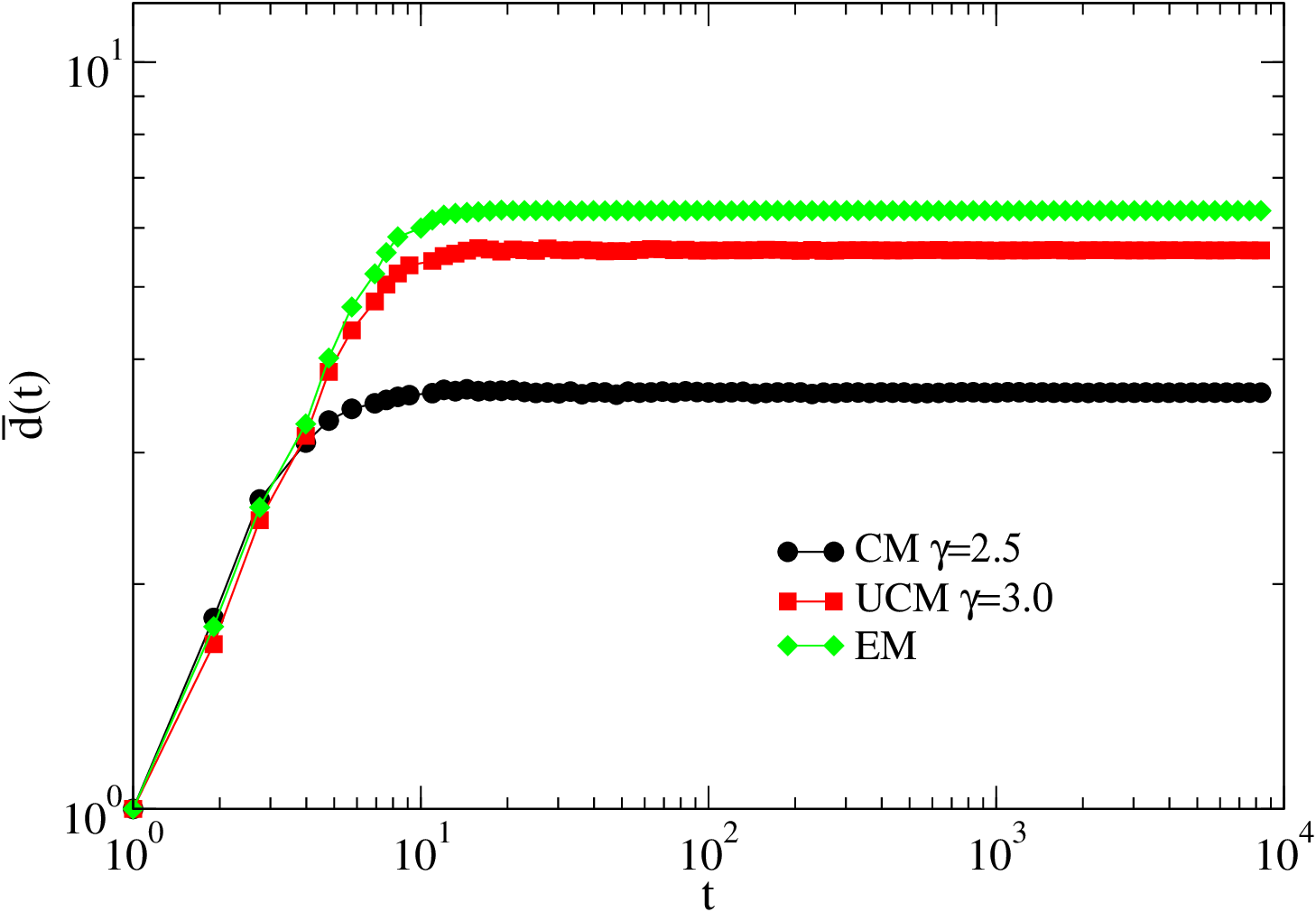}
  }
  \caption{(Color online) MTD as a function of time for looped complex
    networks of size $N=10^6$.}
\label{f:average_distanceL}
\end{figure}

In looped networks, see Fig.~\ref{f:average_distanceL}, we observe a
very rapid increase of the MTD with time.  In previous works
\cite{gallos04:_random_walk}, the growth of the average distance
(there measured instead as the RMSTD) was found to be a power law,
$\bar{d}(t) \sim t^{\alpha}$ at early times in SF networks, with the
exponent $\alpha$ depending on the degree exponent. In our simulations
on looped networks generated with different algorithms, we do not find
a clear signature for a power law behavior, which can only be
approximately found in a tiny range of values of $t$ at the very
beginning of the walk, even for networks of size $N=10^6$.  At large
times, on the other hand, the MTD reaches a plateau, $\bar{d}(\infty)
= \langle \bar{d}\rangle$, due to finite size effects.  The value of
this plateau can be estimated using simple quantitative
arguments. Assume that the random walker starts from a source vertex
of degree $k$. During its dynamics in the steady state, it visits
vertices of degree $k'$ with probability, see Eq.~(\ref{eq:2}), $k' /
\avk N$. On the other hand, vertices of degree $k$ and $k'$ are, in
average, at a topological distance
\cite{holyst05:_univer_scalin,dorogovtsev06:_degree_depen} $d_{k,
  k'}$; therefore, we will expect the random walker to be at an
average distance of a source of degree $k$
\begin{equation}
  \langle \bar{d}\rangle_k = \sum_{k'} \frac{k' P(k')}{\avk} d_{k, k'}. 
\end{equation}
A further average over all possibles sources, leads to an average
distance of the walker, for any source vertex, given by 
\begin{equation}
  \langle \bar{d}\rangle = \sum_k P(k) \langle \bar{d}\rangle_k =
  \sum_{k} \frac{k P(k)}{\avk} d_k,
\end{equation}
where $d_k = \sum_{k'} P(k') d_{k, k'}$ is the mean topological
distance from any vertex to a given vertex of degree $k$
\cite{dorogovtsev06:_degree_depen}. The scaling of $\langle
\bar{d}\rangle$ with system size can be easily predicted assuming the
expressions of $d_k$ in Ref.~\cite{dorogovtsev06:_degree_depen},
namely
\begin{eqnarray}
  d_k &\simeq&  A \ln \left( \frac{N}{ k^{(\gamma-1)/2}} \right) \quad
  \mbox{(SF networks)} \label{eq:6}\\
  d_k &\simeq& A' \ln N - B' k \quad \mbox{(exponential
    networks)}\label{eq:7}, 
\end{eqnarray}
where $A$, $A'$ and $B'$ are size-independent constants. This yields
in both cases
\begin{equation}
  \langle \bar{d}\rangle \simeq \ln N.
\end{equation}

\begin{figure}[t]
  \centerline{
    \includegraphics*[width=0.40\textwidth]{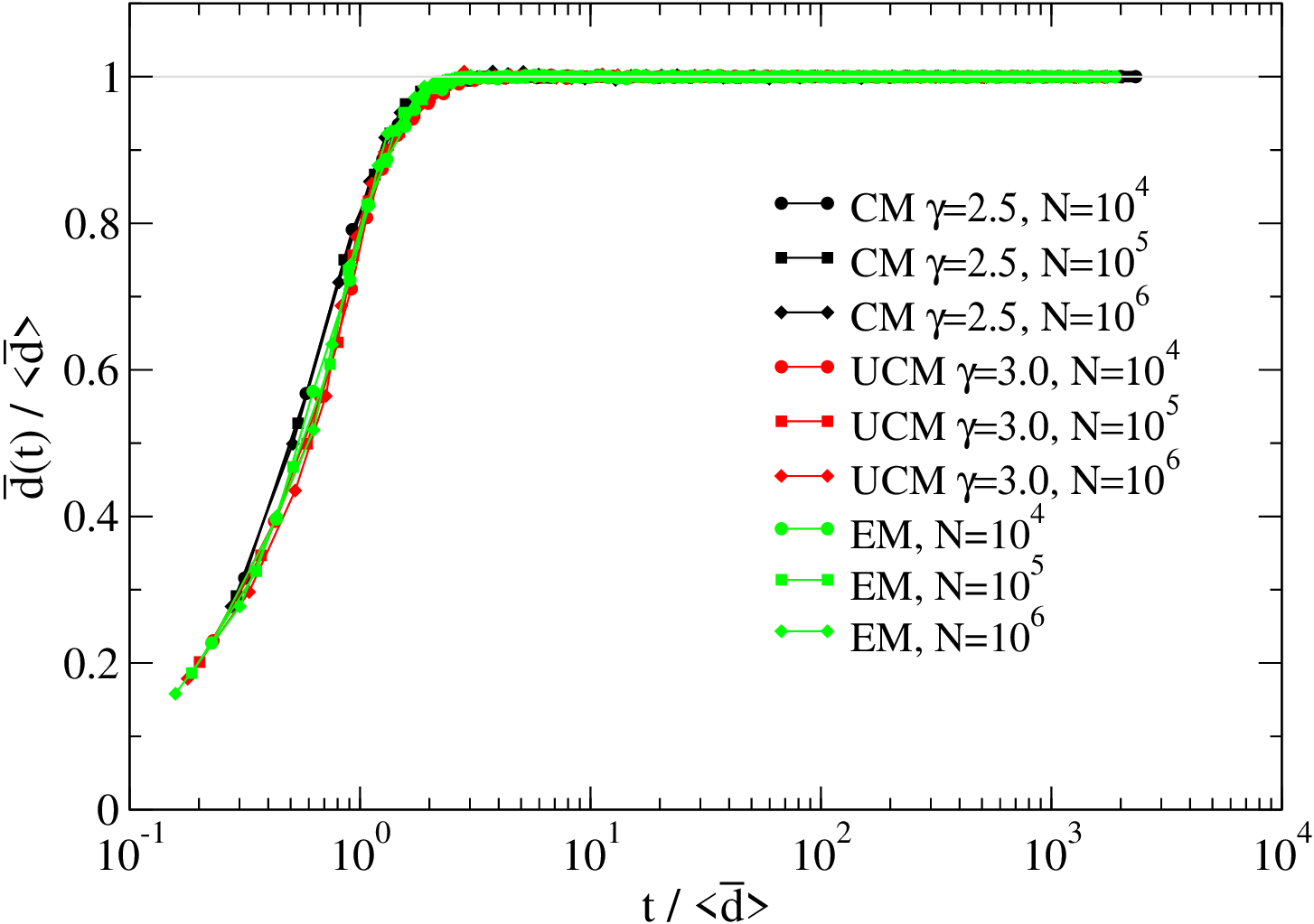}
  }
  \caption{(Color online) Rescaled MTD as a function of time for looped complex
    networks.}
\label{f:average_distanceLScaled}
\end{figure}

Turning to the numerical data for looped networks in
Fig.~\ref{f:average_distanceLScaled}, we observe that it is compatible
with a scaling behavior of the form 
\begin{equation}
  \bar{d}_L(t) = \langle \bar{d} \rangle   
  f \left( \frac{t}{\langle \bar{d} \rangle} \right). 
\end{equation}
This scaling indicates that, after a short characteristic time $t_c
\sim \langle \bar{d} \rangle \sim \ln N$, the walker is in average as far as
the origin at it can be, and it can therefore freely explore the whole
network.

\begin{figure}[t]
  \centerline{
    \includegraphics*[width=0.40\textwidth]{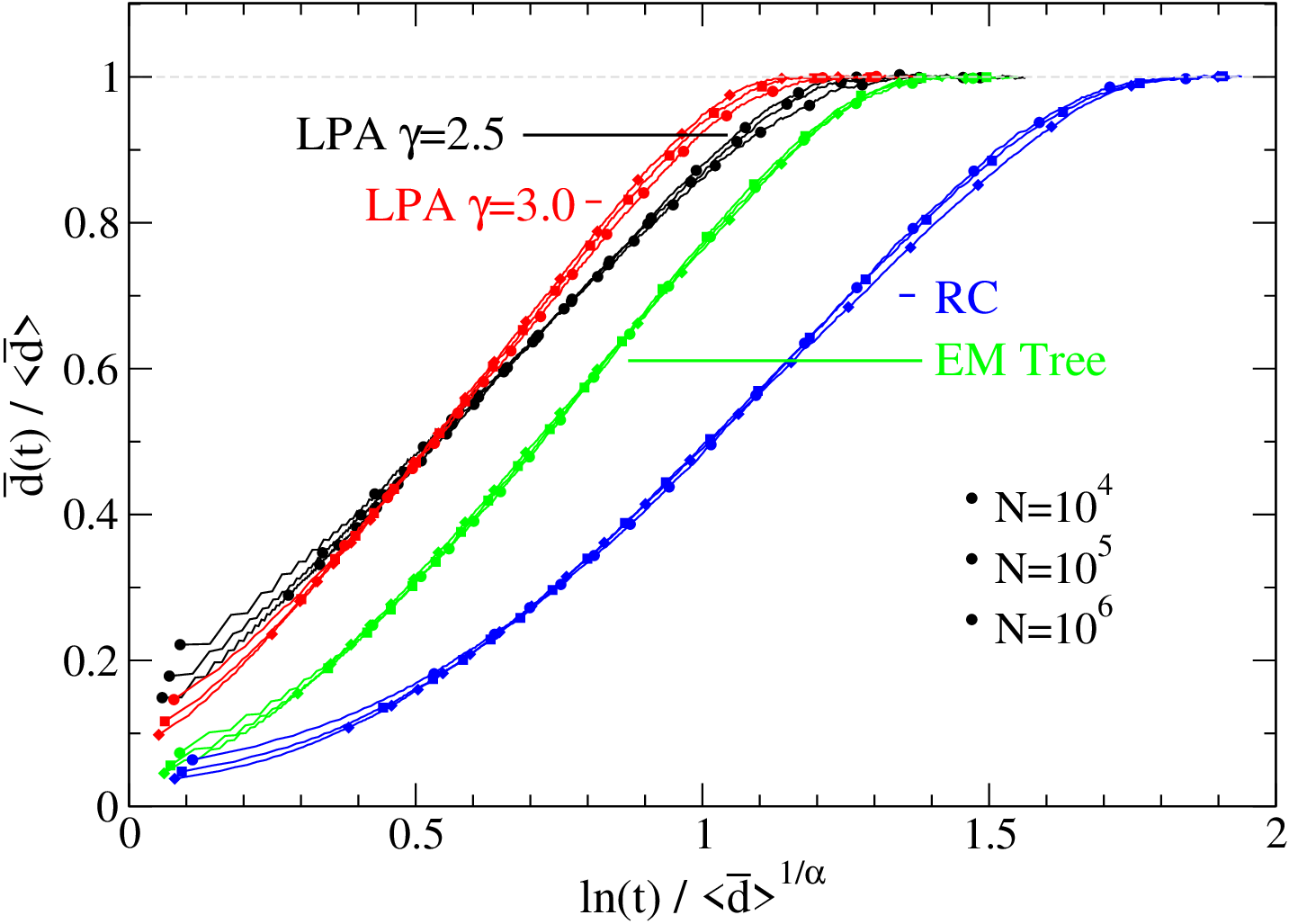}
  }
  \caption{(Color online) MTD as a function of time for complex trees
    of size $N=10^6$. The exponent $\alpha$ is determined by a
    numerical fit from the observed relation $\bar{d}_T(t) \sim (\ln
    t)^\alpha$ (data not shown). The values used in Figure are $\alpha
    \simeq 0.91$ for LPA trees with $\gamma=2.5$, $\alpha \simeq 1.04$
    for LPA trees with $\gamma=3.0$, $\alpha \simeq 1.31$ for EM trees
    and $\alpha \simeq 1.62$ for the RC tree.}
\label{f:average_distanceT}
\end{figure}

In trees, Fig.~\ref{f:average_distanceT}, on the other hand, we
observe a much slower growth of the MTD at early times, which can be
approximately fitted with the form
\begin{equation}
  \bar{d}_T(t)  \sim (\ln t)^\alpha,
\end{equation}
where the exponent $\alpha$ depends on the details of the network. The
whole function $\bar{d}_T(t)$ is also observed to fulfill the scaling form
\begin{equation}
  \bar{d}_T(t) = \langle \bar{d} \rangle   
  f \left( \frac{\ln t}{\langle \bar{d} \rangle^{1 / \alpha}} \right).
\end{equation}
This form implies that the characteristic time to escape from the
neighborhood of the origin scales as $t_c \sim \exp (\langle d
\rangle^{1 / \alpha}) \sim \exp[ (\ln N)^{1 / \alpha}]$, which means
that the exploration process is much more slower in trees, with the
walker spending large amounts of time exploring the close vicinity of
the origin of the walk.  We remark that here the scaling function
$f(x)$ displays some further dependences on degree exponent, average
degree and degree correlations in both looped and tree networks.

\begin{figure}[t]
\centerline{
  \includegraphics*[width=0.40\textwidth]{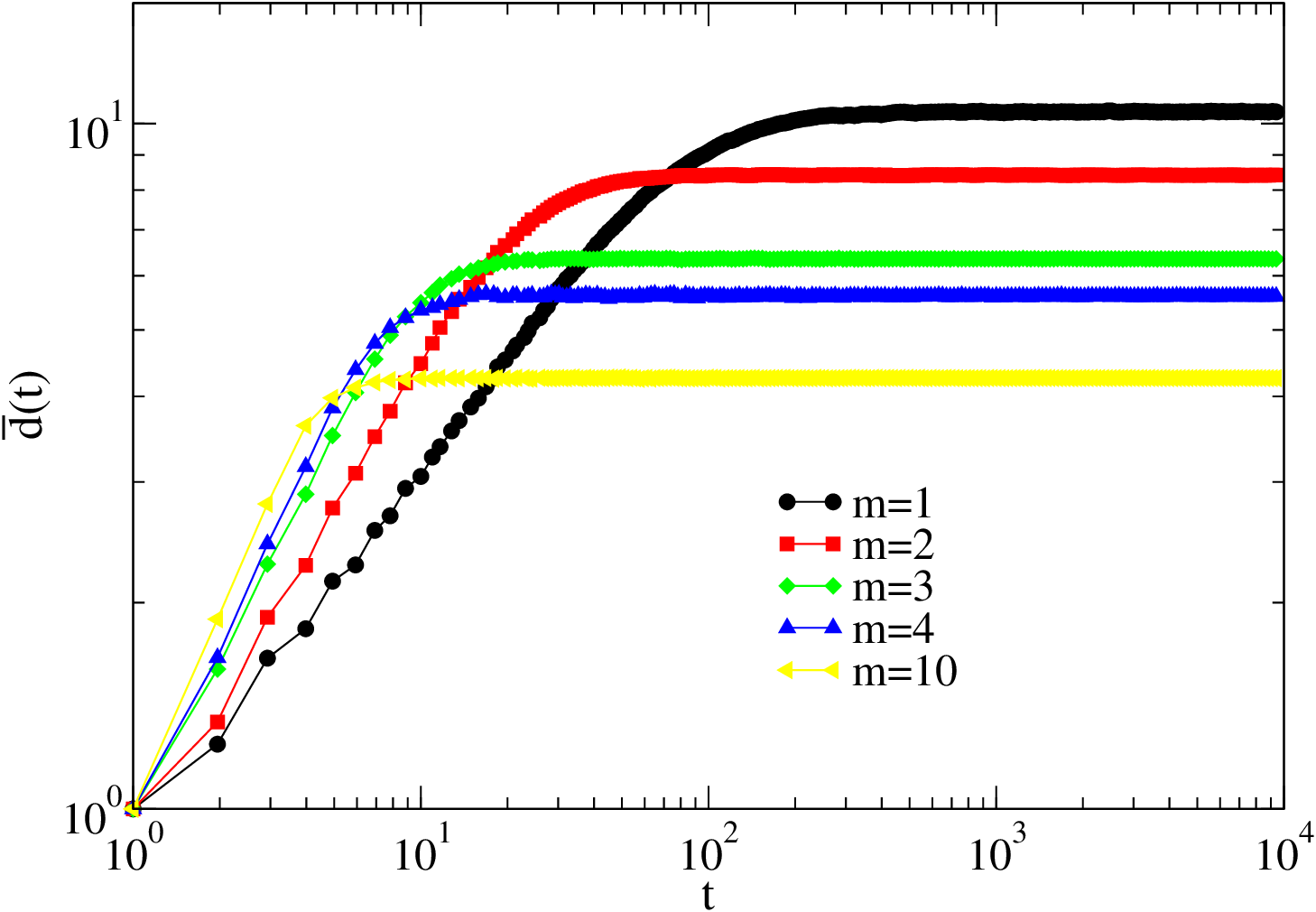}
}	
\caption{(Color online) MTD as a function of time for UCM looped networks
  ($\gamma=3.0$ and $N=10^6$) with varying minimum degree $m$.}
  \label{f:average_distanceDifferentM}
\end{figure}

The fact that the presence of a tree-like structure slows down the
distance explored by a random walker on a network, allows to interpret
the results presented in Ref.~\cite{gallos04:_random_walk}, in
particular the power-law behavior at initial times of $\bar{d}(t)$. In
fact, in Ref.~\cite{gallos04:_random_walk} the substrate for the
random walk simulations were SF networks generated with the CM model
with minimum degree $m=1$. In this case, simulations were performed on
the giant component. Apart from the possible effect of degree
correlations for $\gamma<3$, the point is that, for $m=1$, traces of
tree-like structure are still present in the network, in the form of
chains of small degree vertices \cite{samukhin08:_laplac}.  Thus, a
remnant slowing down effect of the tree component is observed, see
Fig.~\ref{f:average_distanceDifferentM}, leading to an MTD that, at
short times, scales as $\bar{d}(t) \sim t^{0.55}$ for the data at $m=1$
shown in this graph, in excellent agreement with the observation in
\cite{gallos04:_random_walk}, namely $\bar{d^2}(t)\sim t^{1.1}$ for
the RMSTD.

A further remark concerns the relation between our results and the
above mentioned analytical calculations for Bethe
lattices~\cite{hughes82}, according to which these structures exhibit
a behavior analogous to the one observed in looped networks. The
apparent incongruity vanishes when noticing that, while Bethe lattices
are infinite hierarchical structures, we have focused on complex
(i.e. disordered) finite trees. To recover numerically the Bethe
lattice behavior, indeed, it is necessary to adopt special algorithms
in order to simulate an infinite hierarchical
tree~\cite{argyrakis2000rwa,katsoulis2002dat}.

\section{Mean first-passage time}
\label{sec:mean-first-passage}

More information about the dynamics of random walks can be extracted
from the analysis of the mean first passage time (MFPT)
\cite{redner_fpp} $\tau(i\to j)$, defined as the average time that a
random walker takes to arrive for the first time at vertex $j$,
starting from vertex $i$ \cite{nohrieger}. In networks with no
translation symmetry, the MFPT from a source $i$ to a target $j$ needs
not be equal to the MFTP from source $j$ to target $i$. Therefore,
different reduced MFPTs can be considered. We can thus define the
direct MFTP $\tau^\rightarrow(k)$ as the MFPT on a target vertex of
degree $k$, starting from a randomly chosen source vertex, and the
inverse MFPT $\tau^\leftarrow(k)$ as the MFTP on a randomly target
vertex, starting from a source vertex of degree $k$, namely
\begin{eqnarray}
  \tau^\rightarrow(k) &=&  \frac{1}{N} \sum_i \frac{1}{N_k} \sum_{j\in
    \mathcal{V}(k)} \tau(i\to j),\label{eq:12}\\
  \tau^\leftarrow(k)  &=&\frac{1}{N} \sum_j \frac{1}{N_k} \sum_{i\in
    \mathcal{V}(k)} \tau(i\to j),\label{eq:13}
\end{eqnarray}
where $\mathcal{V}(k)$ is the set of vertices of degree $k$ and $N_k$
is the number of such vertices.  A simple argument can predict the
form of the MFPTs for random uncorrelated networks.  In this case, the
probability for the walker to arrive at a vertex $i$, in a hop
following a randomly chosen edge, is given by $q(i)=q(k_i)=k_i/\avk N$
\cite{newmanrev}.  Therefore, the probability of arriving at vertex
$i$ for the first time after $t$ hops is $P_a(i;t) = [1-q(i)]^{t-1}
q(i)$. The direct MFTP to vertex $i$ can thus be estimated as the
average
\begin{equation}
  \tau^\rightarrow(k_i) = \sum_t t P_a(i;t) = \frac{\avk N}{k_i}.
\end{equation}
For the inverse MFPT, we notice that, in a random network, after the
first hop, the walker loses completely the memory of its source
degree, therefore we can approximate 
\begin{equation}
  \tau^\leftarrow(k) = \sum_k P(k) \tau^\rightarrow(k) = \avk \langle
  k^{-1} \rangle N.
\end{equation}

\begin{figure}[t]
  \centerline{
  \includegraphics*[width=0.40\textwidth]{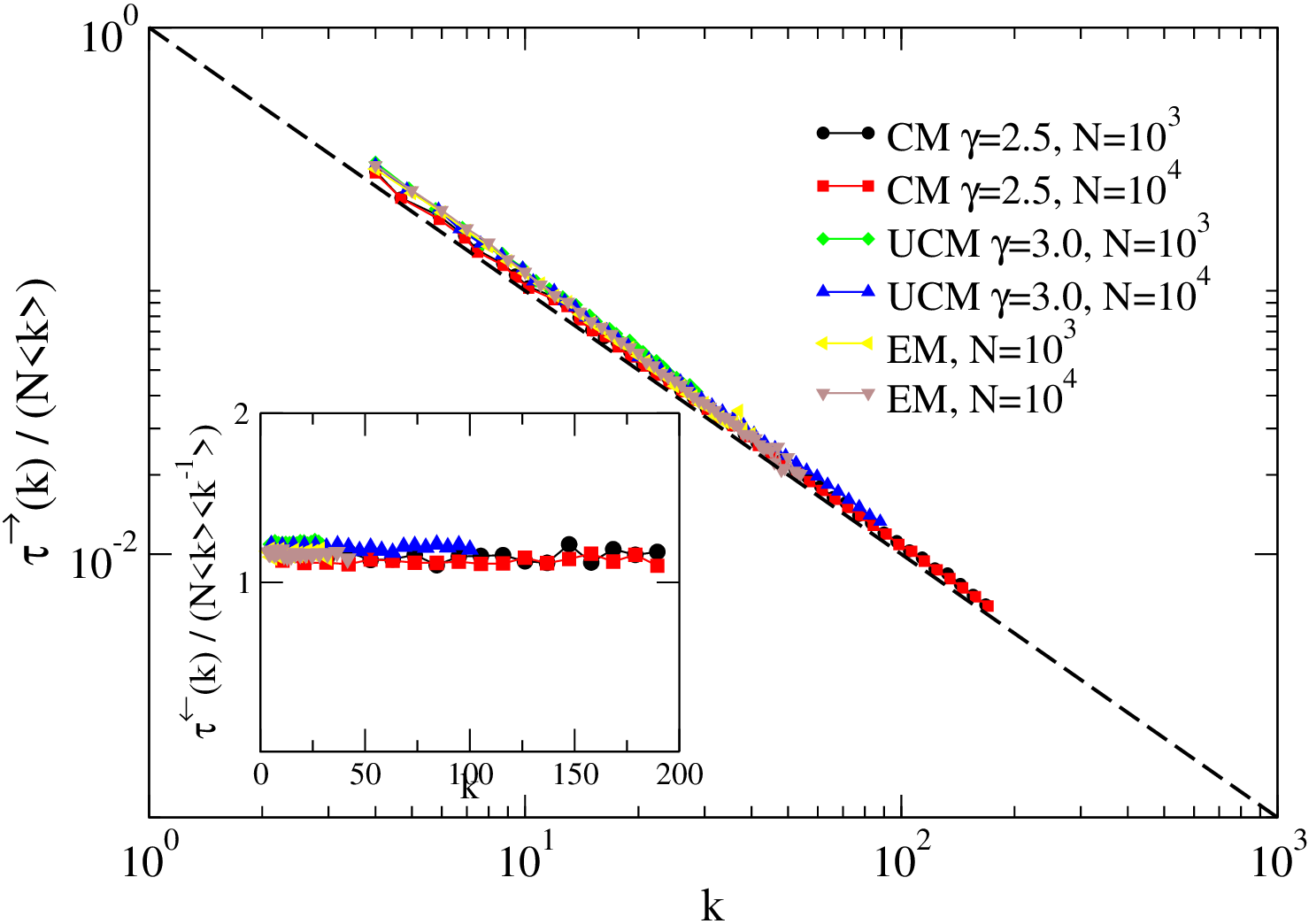}
}	
  \caption{(Color online) Reduced MFPTs as a function of the degree $k$ for looped
    complex networks. We recover the
    simple mean-field predictions $\tau^\rightarrow_L(k) \simeq \avk N
    /k$ (main figure) and $\tau^\leftarrow_L(k) \simeq \avk \langle k^{-1} \rangle
    N$ (inset).}
\label{f:mfpt_spectraL}
\end{figure}

Less trivial approaches~\cite{nohrieger,baronchelli_rw2006} show in
fact that the MFPT from a source vertex $i$ to target vertex $j$
depends on the degree of the target vertex as $\tau(i\to j) \sim
1/k_j$, but has a residual dependence on the source vertex and it is
actually asymmetric, $\tau(i\to j) \neq \tau( j \to i)$. This fact
could in principle affect the form of the reduced MFPTs in real
networks, defined in Eqs.~(\ref{eq:12}) and~(\ref{eq:13}).
Fig.~\ref{f:mfpt_spectraL}, however, shows that for looped networks
the behavior predicted for random uncorrelated networks turns out to
be extremely robust with respect to changes in the topological
properties of the network: homogeneous or heterogenous nature, degree
exponent, presence or absence of correlations,
etc. \cite{structurednets,gallos04:_random_walk,baronchelli_rw2006}.

\begin{figure}[t]
  \centerline{
    \includegraphics*[width=0.40\textwidth]{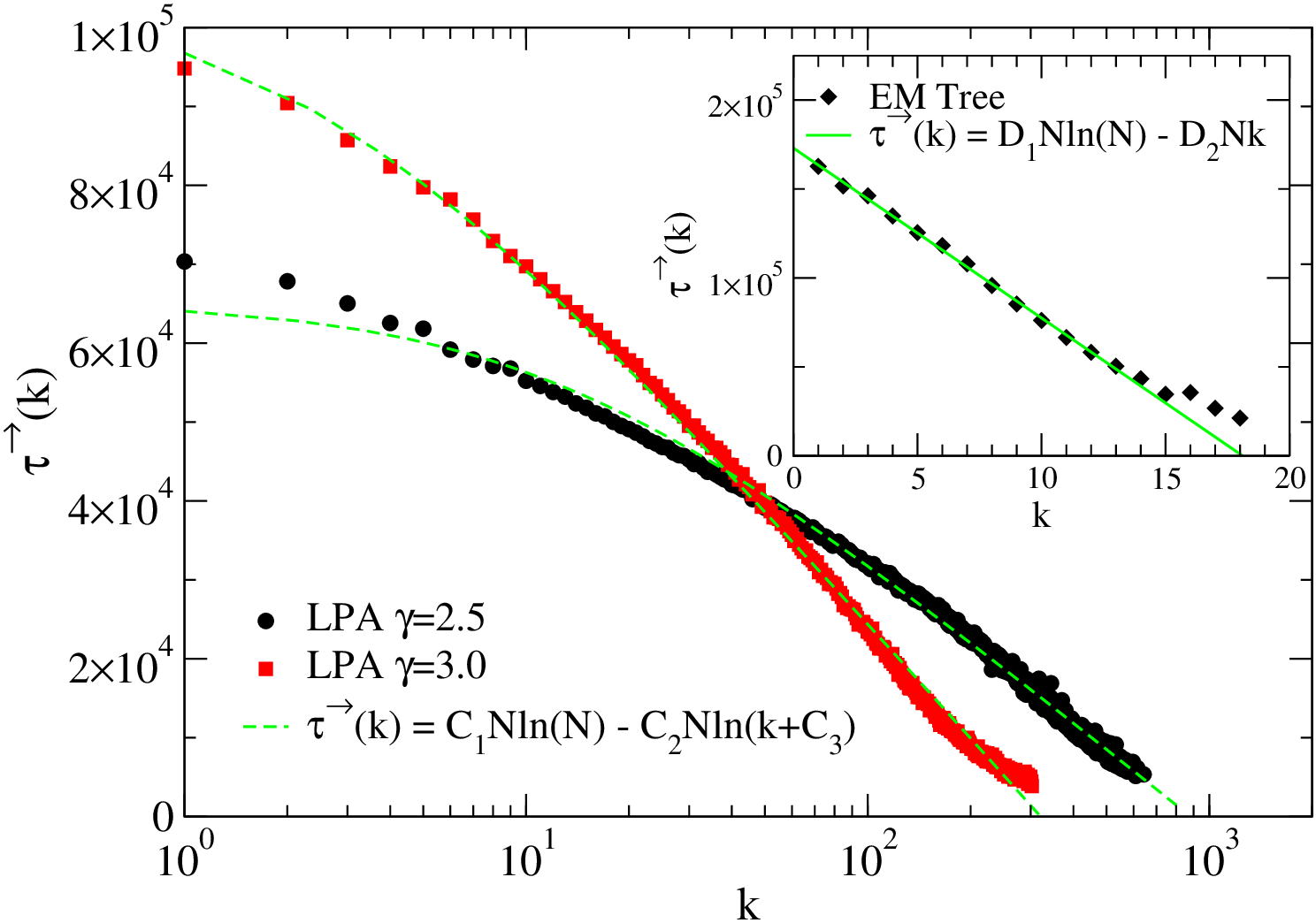}
  }	
  \caption{(Color online) Direct MFPT as a function of the degree $k$
    for SF tree networks ($N=10^4$). Dashed lines correspond to
    nonlinear fittings to the empirical form Eq.~(\ref{eq:11}). Inset:
    Direct MFPT as a function of the degree $k$ for homogeneous EM
    tree networks. The dashed line corresponds to a fitting to the
    empirical form Eq.~(\ref{eq:14}).}
  \label{f:mfpt_spectraT}
\end{figure}

\begin{figure}[t]
  \centerline{
  \includegraphics*[width=0.40\textwidth]{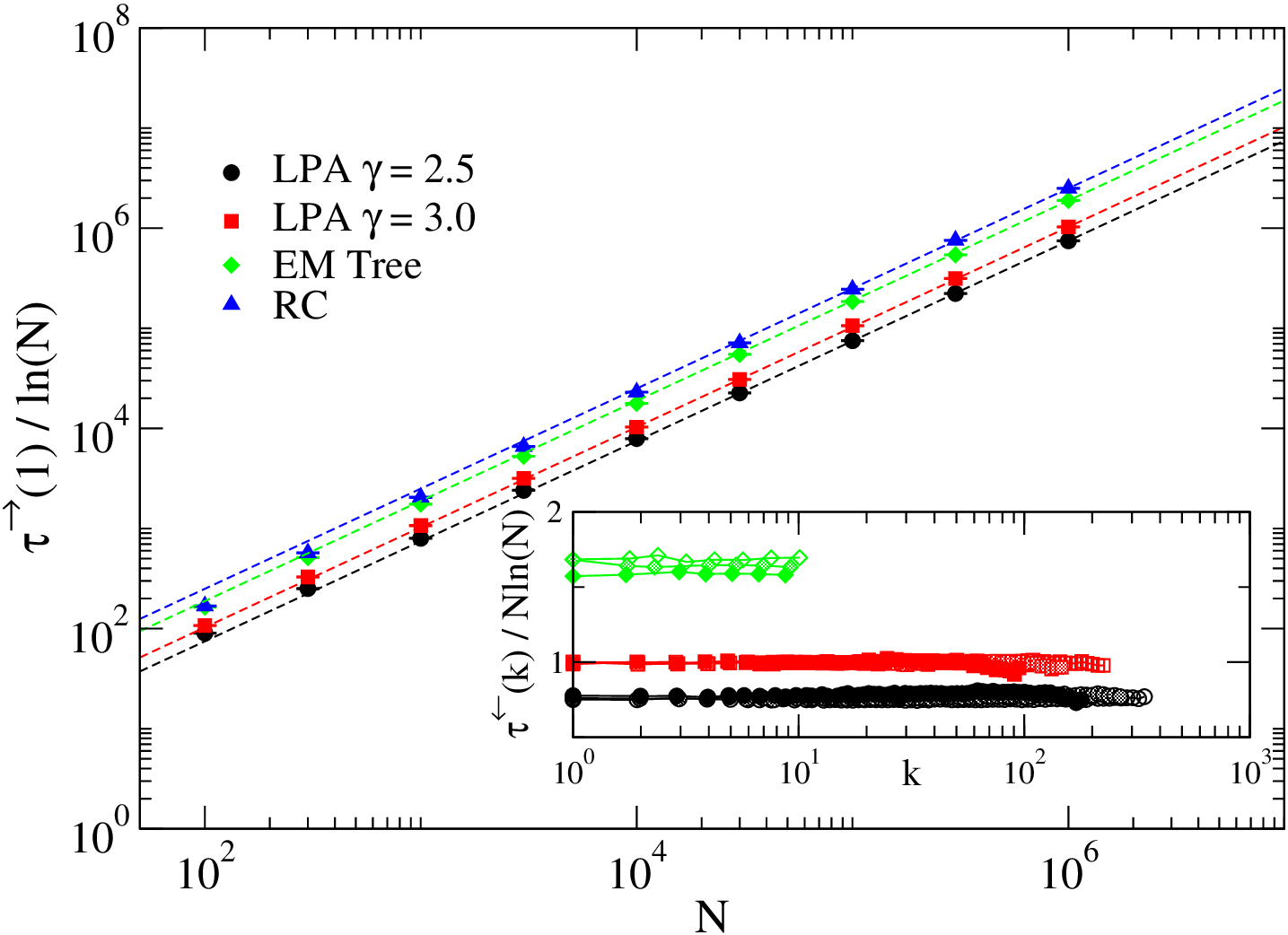}
}	
\caption{(Color online) Direct MFPT on leaves in complex trees as a function of the
  network size $N$.  The observed scaling is $\tau^\rightarrow_T(1)
  \sim N \ln N$. Inset: Inverse MFPT on complex trees for different
  network sizes ($N=10^3$ full colored points, $N=3 \times 10^3$ light colored 
  points, $N=10^4$ empty points). The observed scaling is again $\tau^\leftarrow_T(k)
  \sim N \ln N$.}
\label{f:mfpt_on_k1}
\end{figure}

In trees, on the other hand, we find a completely different picture,
see Fig.~\ref{f:mfpt_spectraT}. Now, the direct MFPT in SF trees
decays with $k$ much slower than in looped networks.  In fact, we can
fit it numerically to the form
\begin{equation}
  \tau^\rightarrow_T(k) = C_1 N \ln N - C_2 N \ln (k+C_3),
  \label{eq:11}
\end{equation}
where $C_1$, $C_2$ and $C_3$ fitting parameters that depend only
slightly on the network size. The $N \ln N$ dependence can be directly
observed by plotting $\tau^\rightarrow_T(1)$ for different system
sizes, as shown in Fig.~\ref{f:mfpt_on_k1}. For homogeneous EM
networks, on the other hand, the direct MFPT can be fitted to the form
\begin{equation}
  \tau^\rightarrow_T(k) = D_1 N \ln N - D_2 N k,
  \label{eq:14}
\end{equation}
see inset in Fig.~\ref{f:mfpt_spectraT}.  The scaling of
$\tau^\rightarrow_T(1)$ in this case is also checked in
Fig.~\ref{f:mfpt_on_k1}.  With respect to the inverse MFTP, it is
again constant, but now scales with system size as
$\tau^\leftarrow_T(k) \sim N \ln N$ for all kinds of trees (inset in
Fig.~\ref{f:mfpt_on_k1}).

The topological structure of the trees can explain the unusual form of
the MFPTs. While in looped networks the number of access paths to the
target vertex is related to its degree, on the tree the path is
unique, and is given by the one-dimensional set of links and vertices
connecting the starting vertex to the target. In this case, the degree
of the target is much less important from the point of view of the
walker, since finding the target corresponds to finding a particular
\textit{leaf} (i.e. a $k=1$ vertex) of the sub-tree the random walker
is exploring.  This observation suggests that while in looped networks
the MFPT into a vertex is dominated by its degree (because the latter
is related the multiplicity of the entry paths to the vertex), in
trees the distance between the source and the target can be much more
relevant, and thus induce a larger MFPT.


\begin{figure}[t]
\centerline{
\includegraphics*[width=0.40\textwidth]{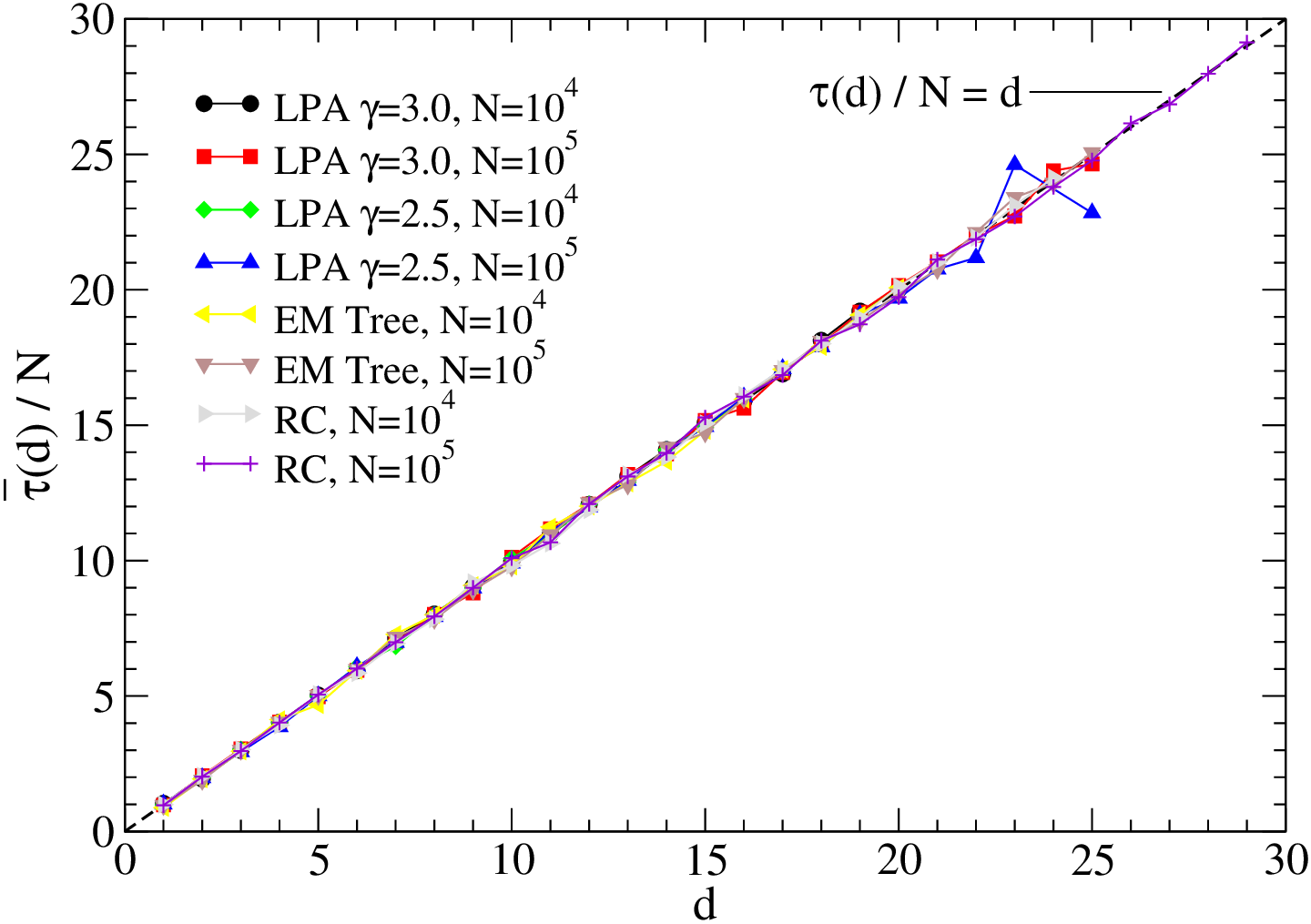}
}
\caption{(Color online) MRTT $\bar{\tau}(d)$ as a function of the source-target
  topological distance $d$ in trees. Different curves collapse
  perfectly on $\bar{\tau}_T(d) \sim Nd$.}
\label{f:mfpt_vs_d}
\end{figure}

We therefore consider the MFPT as a function of the topological
distance $d_{ij}$ between the starting vertex $i$ and the target
$j$~\cite{baronchelli_rw2006,condamin_nature}. 
Since the distance between two vertices is by definition a symmetric quantity,
it seems natural to re-define the MFPT in terms of the symmetric
mean round trip time (MRTT)
\begin{equation}
  \bar{\tau}(d_{ij}) = \tau(i \to j) + \tau(j \to i),
\end{equation}
i.e. the average time to go from $i$ to $j$ and back or vice-versa. It
has been recently proved~\cite{condamin_nature} that, for complex
scale-invariant networks, the MRTT averaged for all vertices at the
same distance scales as
\begin{equation}
  \bar{\tau}(d) \simeq  N d^{D_w - D_b},
  \label{eq:5}
\end{equation}
where $D_b$ is the box dimension of the network, and $D_w$ its walk
exponent \cite{song2006ofg}. For a class of scale-invariant networks
\cite{song2006ofg} corresponding to a tree structure, for which $D_w -
D_b=1$, the authors of Ref.~\cite{condamin_nature} obtained
correspondingly a linear scaling $\bar{\tau}_T(d) \simeq Nd$.  We have
checked that this linear form holds for different SF, EM and RC trees,
see Fig.~\ref{f:mfpt_vs_d}, a result that leads us to conjecture that,
for any complex tree, $D_w - D_b=1$.

\begin{figure}[t]
\centerline{
\includegraphics*[width=0.4\textwidth]{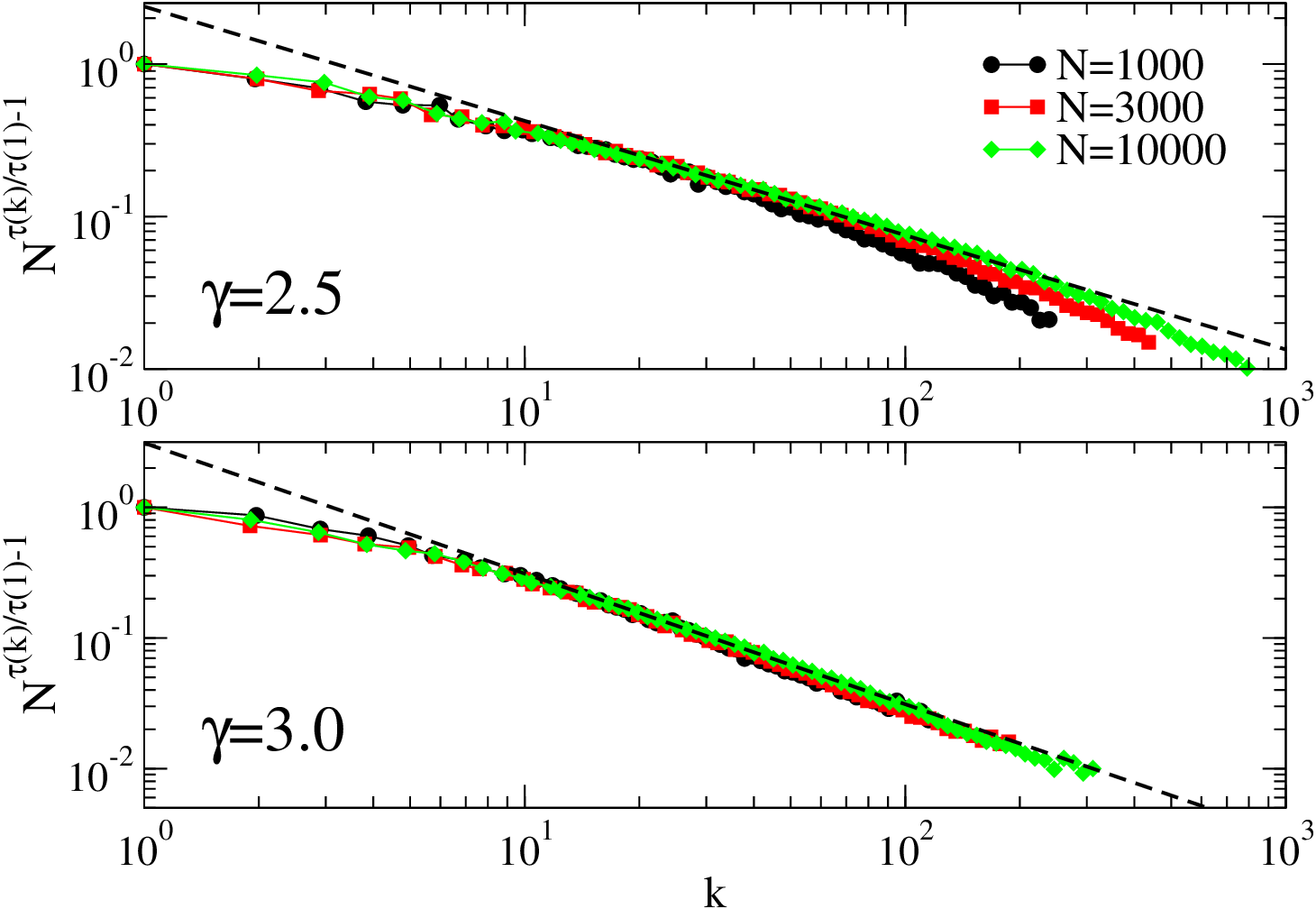}
}
\caption{(Color online) Rescaled MRTT $\bar{\tau}_T(t)$ as a function
  of the source degree $k$ and for randomly chosen targets in SF
  trees. Predictions of Eq.~(\ref{e:rescaling}), i.e.  $N^{\tau(k) /
    \tau(1)-1} \sim k^{(1-\gamma)/2}$, are plotted as dashed lines.}
\label{f:mfpt_k_rescaled}
\end{figure}

We can use the result in Eq.~(\ref{eq:5}) to gain insight on the
behavior of the anomalous reduced MFPTs in tree networks. Considering
an average over all vertices with the same degree, we have that
$\bar{\tau}(d_{k k'}) = \tau(k \to k') + \tau(k' \to k)$. Averaging
now over $k'$, we can consider the reduced MRTT
\begin{equation}
  \bar{\tau}(k) = \sum_{k'} P(k') [ \tau(k \to k') + \tau(k' \to k)] =
   \tau^\leftarrow(k) + \tau^\rightarrow(k),
  \label{eq:10}
\end{equation}
defined as the average time to go from a randomly chosen vertex to a
given vertex of degree $k$, and back (or vice-versa, since the MRTT is
symmetric).  Now, since $\bar{\tau}(d_{k k'})$ is linear in $d_{k k'}$
for tree networks, we have
\begin{equation}
   \bar{\tau}(k) \simeq  \sum_{k'} P(k') N d_{k k'} = N d_{k}.
\end{equation}
Assuming the scaling of $d_k$ as given by Eqs.~(\ref{eq:6})
and~(\ref{eq:7}), we obtain 
\begin{equation}
  \bar{\tau}_T(k) \simeq N A \ln \left(
    \frac{N}{k^{(\gamma-1)/2}}\right)  
  \label{eq:8}
\end{equation}
for SF networks and
\begin{equation}
  \bar{\tau}_T(k) \simeq N  \left(
    A' \ln N - B' k \right)  
  \label{eq:9}
\end{equation}
for EM networks.  The unknown constant in Eq.~(\ref{eq:8})
can be reabsorbed in the the value of $\bar{\tau}_T(1)$, to obtain
a scaling form with system size for SF networks that reads
\begin{equation}
  \frac{\bar{\tau}_T(k)}{\bar{\tau}_T(1)} \sim \frac{1}{\ln N} \ln
  \left( 
    \frac{N}{k^{(\gamma-1)/2}} \right).
\label{e:rescaling}
\end{equation}
In Fig.~\ref{f:mfpt_k_rescaled} we show that this scaling form is very
well satisfied by the MRTT in SF trees, independently of the degree
exponent and correlation patterns, at least for intermediate values of
$k$. The observed bending at small degrees can be ascribed to the
presence of a constant in the logarithm analogous to empirical
parameter $C_3$ in Eq.~(\ref{eq:11}), that does not follow from our
argument. Finite size effects, on the other hand, are responsible for
the deviations present at large degrees, that are indeed more evident
in SF trees with smaller values of $\gamma$.

This observations allow us to interpret the anomalous functional form
of the reduced MFTPs observed in trees. From Eq.~(\ref{eq:10}), we
have
\begin{equation}
  \tau^\rightarrow_T(k) = \bar{\tau}_T(k) -  \tau^\leftarrow_T(k).
\end{equation}
Writing $\tau^\leftarrow_T(k) \sim C N \ln N$, from Eq.~(\ref{eq:8}) we
obtain, for SF networks,
\begin{equation}
  \tau^\rightarrow_T(k) \sim  (A-C) N \ln N
  - \frac{A(\gamma-1)}{2} N \ln k 
\end{equation}
while for homogeneous EM networks, we have
\begin{equation}
  \tau^\rightarrow_T(k) \sim  (A'-C) N \ln N 
  - B' N k,
\end{equation} 
in agreement with the empirical fitting found in Eqs.~(\ref{eq:11})
and~(\ref{eq:14}).

This argument cannot be extended to looped networks, since here
$\bar{\tau}(d_{k k'})$ is not linear in $d_{k k'}$.  The $k$
dependence of the MRTT can be however trivially obtained from the
reduced MFPTs as
$\bar{\tau}_L(k)=\tau^\rightarrow_L(k)+\tau^\leftarrow_L(k)\simeq \avk
N(\langle k^{-1} \rangle + 1/k)$.

\section{Conclusions}

In this paper we have shown that complex
tree-like topologies heavily affect the behavior of a random walk
performed on top of them, with a global slowing down of the dynamics
and a logarithmic dependence of the first passage time properties in
SF networks. These features are intrinsically connected with the
complex tree structure and cannot be attributed to the mere presence
of leaves, while they are radically different from the ones exhibited
by Bethe lattices, i.e. infinite and hierarchical tree structures.

We have studied the random walk exploration properties and we have
shown that complex trees induce a slower dynamics, compared to looped
networks, for both the coverage and the mean topological displacement
problems.  Moreover, by means of the analysis of the symmetrized MFPT
(the MRTT), we have been able to recognize the different role played
by the degree $k$ of the target vertex in looped and tree
structures. In the former, a larger degree corresponds to a larger
number of access ways to the target vertex. In the latter, on the
other hand, the target vertex is always seen as a leaf by the random
walker, and its degree $k$ affects the MFPT only through the
dependence of the average distance $d_k$ between it and the rest of
the vertices. These results provide important insights into diffusion
problems on trees, and help explaining the characteristic slow
dynamics observed on diffusive processes taking place on top of tree
networks~\cite{castellano_voter2,dallasta_ng_nets,nohkim}. Moreover,
they are also interesting in the study of dynamics in real-world
networks, in which the so-called border trees motifs
~\cite{villasboas2007btc} have been recently shown to be significantly
present.

\section*{Acknowledgments}

We acknowledge financial support from the Spanish MEC (FEDER), under
project No. FIS2007-66485-C02-01, and additional support from the
DURSI, Generalitat de Catalunya (Spain). M. C. acknowledges financial
support from Universitat Polit\`ecnica de Catalunya.


\begin{thebibliography}{42}
\expandafter\ifx\csname natexlab\endcsname\relax\def\natexlab#1{#1}\fi
\expandafter\ifx\csname bibnamefont\endcsname\relax
  \def\bibnamefont#1{#1}\fi
\expandafter\ifx\csname bibfnamefont\endcsname\relax
  \def\bibfnamefont#1{#1}\fi
\expandafter\ifx\csname citenamefont\endcsname\relax
  \def\citenamefont#1{#1}\fi
\expandafter\ifx\csname url\endcsname\relax
  \def\url#1{\texttt{#1}}\fi
\expandafter\ifx\csname urlprefix\endcsname\relax\def\urlprefix{URL }\fi
\providecommand{\bibinfo}[2]{#2}
\providecommand{\eprint}[2][]{\url{#2}}

\bibitem[{\citenamefont{Huberman and Kerszberg}(1985)}]{huberman1985}
\bibinfo{author}{\bibfnamefont{B.}~\bibnamefont{Huberman}} \bibnamefont{and}
  \bibinfo{author}{\bibfnamefont{M.}~\bibnamefont{Kerszberg}},
  \bibinfo{journal}{J. Phys. A: Math. Gen.} \textbf{\bibinfo{volume}{18}},
  \bibinfo{pages}{L331} (\bibinfo{year}{1985}).

\bibitem[{\citenamefont{Bachas and Huberman}(1986)}]{huberman1986}
\bibinfo{author}{\bibfnamefont{C.~P.} \bibnamefont{Bachas}} \bibnamefont{and}
  \bibinfo{author}{\bibfnamefont{B.~A.} \bibnamefont{Huberman}},
  \bibinfo{journal}{Phys. Rev. Lett.} \textbf{\bibinfo{volume}{57}},
  \bibinfo{pages}{1965} (\bibinfo{year}{1986}).

\bibitem[{\citenamefont{Sibani and Hoffmann}(1989)}]{sibani1989}
\bibinfo{author}{\bibfnamefont{P.}~\bibnamefont{Sibani}} \bibnamefont{and}
  \bibinfo{author}{\bibfnamefont{K.~H.} \bibnamefont{Hoffmann}},
  \bibinfo{journal}{Phys. Rev. Lett.} \textbf{\bibinfo{volume}{63}},
  \bibinfo{pages}{2853} (\bibinfo{year}{1989}).

\bibitem[{\citenamefont{Lee and Raymond}(1993)}]{lee_microcomputer}
\bibinfo{author}{\bibfnamefont{E.}~\bibnamefont{Lee}} \bibnamefont{and}
  \bibinfo{author}{\bibfnamefont{D.}~\bibnamefont{Raymond}},
  \bibinfo{journal}{Encyclopedia of Microcomputers}
  \textbf{\bibinfo{volume}{11}}, \bibinfo{pages}{101} (\bibinfo{year}{1993}).

\bibitem[{\citenamefont{Card et~al.}(1999)\citenamefont{Card, Mackinlay, and
  Schneiderman}}]{card_book}
\bibinfo{author}{\bibfnamefont{S.}~\bibnamefont{Card}},
  \bibinfo{author}{\bibfnamefont{J.}~\bibnamefont{Mackinlay}},
  \bibnamefont{and}
  \bibinfo{author}{\bibfnamefont{B.}~\bibnamefont{Schneiderman}},
  \emph{\bibinfo{title}{{Readings in Information Visualization: Using Vision to
  Think}}} (\bibinfo{publisher}{Morgan Kaufmann}, \bibinfo{address}{San
  Francisco}, \bibinfo{year}{1999}).

\bibitem[{\citenamefont{Cavalli-Sforza and Edwards}(1967)}]{cavallisforza1967}
\bibinfo{author}{\bibfnamefont{L.}~\bibnamefont{Cavalli-Sforza}}
  \bibnamefont{and} \bibinfo{author}{\bibfnamefont{A.}~\bibnamefont{Edwards}},
  \bibinfo{journal}{Evolution} \textbf{\bibinfo{volume}{21}},
  \bibinfo{pages}{550} (\bibinfo{year}{1967}).

\bibitem[{\citenamefont{Fisher}(1987)}]{fisher_machine_learning}
\bibinfo{author}{\bibfnamefont{D.}~\bibnamefont{Fisher}},
  \bibinfo{journal}{Machine Learning} \textbf{\bibinfo{volume}{2}},
  \bibinfo{pages}{139} (\bibinfo{year}{1987}).

\bibitem[{\citenamefont{Albert and Barab{\'a}si}(2002)}]{barabasi02}
\bibinfo{author}{\bibfnamefont{R.}~\bibnamefont{Albert}} \bibnamefont{and}
  \bibinfo{author}{\bibfnamefont{A.-L.} \bibnamefont{Barab{\'a}si}},
  \bibinfo{journal}{Rev. Mod. Phys.} \textbf{\bibinfo{volume}{74}},
  \bibinfo{pages}{559} (\bibinfo{year}{2002}).

\bibitem[{\citenamefont{Dorogovtsev and Mendes}(2003)}]{mendesbook}
\bibinfo{author}{\bibfnamefont{S.~N.} \bibnamefont{Dorogovtsev}}
  \bibnamefont{and} \bibinfo{author}{\bibfnamefont{J.~F.~F.}
  \bibnamefont{Mendes}}, \emph{\bibinfo{title}{Evolution of networks: From
  biological nets to the {I}nternet and {WWW}}} (\bibinfo{publisher}{Oxford
  University Press}, \bibinfo{address}{Oxford}, \bibinfo{year}{2003}).

\bibitem[{\citenamefont{Dorogovtsev et~al.}(2007)\citenamefont{Dorogovtsev,
  Goltsev, and Mendes}}]{dorogovtsev_critical2007}
\bibinfo{author}{\bibfnamefont{S.}~\bibnamefont{Dorogovtsev}},
  \bibinfo{author}{\bibfnamefont{A.}~\bibnamefont{Goltsev}}, \bibnamefont{and}
  \bibinfo{author}{\bibfnamefont{J.}~\bibnamefont{Mendes}},
  \emph{\bibinfo{title}{Critical phenomena in complex networks}}
  (\bibinfo{year}{2007}), \bibinfo{note}{e-print arXiv:0705.0010v2}.

\bibitem[{\citenamefont{Barab{\'a}si and Albert}(1999)}]{barabasi_albert_first}
\bibinfo{author}{\bibfnamefont{A.-L.} \bibnamefont{Barab{\'a}si}}
  \bibnamefont{and} \bibinfo{author}{\bibfnamefont{R.}~\bibnamefont{Albert}},
  \bibinfo{journal}{Science} \textbf{\bibinfo{volume}{286}},
  \bibinfo{pages}{509} (\bibinfo{year}{1999}).

\bibitem[{\citenamefont{{C. Castellano et al.}}(2005)}]{castellano_voter2}
\bibinfo{author}{\bibnamefont{{C. Castellano et al.}}}, \bibinfo{journal}{Phys.
  Rev. E} \textbf{\bibinfo{volume}{71}}, \bibinfo{pages}{066107}
  (\bibinfo{year}{2005}).

\bibitem[{\citenamefont{{Dall'Asta} et~al.}(2006)\citenamefont{{Dall'Asta},
  Baronchelli, Barrat, and Loreto}}]{dallasta_ng_nets}
\bibinfo{author}{\bibfnamefont{L.}~\bibnamefont{{Dall'Asta}}},
  \bibinfo{author}{\bibfnamefont{A.}~\bibnamefont{Baronchelli}},
  \bibinfo{author}{\bibfnamefont{A.}~\bibnamefont{Barrat}}, \bibnamefont{and}
  \bibinfo{author}{\bibfnamefont{V.}~\bibnamefont{Loreto}},
  \bibinfo{journal}{Phys. Rev. E} \textbf{\bibinfo{volume}{74}},
  \bibinfo{pages}{036105} (\bibinfo{year}{2006}).

\bibitem[{\citenamefont{Noh and Kim}(2006)}]{nohkim}
\bibinfo{author}{\bibfnamefont{J.~D.} \bibnamefont{Noh}} \bibnamefont{and}
  \bibinfo{author}{\bibfnamefont{S.~W.} \bibnamefont{Kim}},
  \bibinfo{journal}{Journal of the Korean Physical Society}
  \textbf{\bibinfo{volume}{48}}, \bibinfo{pages}{S202} (\bibinfo{year}{2006}).

\bibitem[{\citenamefont{Nakamaru and Levin}(2004)}]{nakamaru2004stl}
\bibinfo{author}{\bibfnamefont{M.}~\bibnamefont{Nakamaru}} \bibnamefont{and}
  \bibinfo{author}{\bibfnamefont{S.}~\bibnamefont{Levin}},
  \bibinfo{journal}{Journal of Theoretical Biology}
  \textbf{\bibinfo{volume}{230}}, \bibinfo{pages}{57} (\bibinfo{year}{2004}).

\bibitem[{\citenamefont{Serrano et~al.}(2007)\citenamefont{Serrano,
  Bogu{\~n}{\'a}, Pastor-Satorras, and Vespignani}}]{serrano07:_correl}
\bibinfo{author}{\bibfnamefont{M.~A.} \bibnamefont{Serrano}},
  \bibinfo{author}{\bibfnamefont{M.}~\bibnamefont{Bogu{\~n}{\'a}}},
  \bibinfo{author}{\bibfnamefont{R.}~\bibnamefont{Pastor-Satorras}},
  \bibnamefont{and}
  \bibinfo{author}{\bibfnamefont{A.}~\bibnamefont{Vespignani}}, in
  \emph{\bibinfo{booktitle}{Large scale structure and dynamics of complex
  networks: From information technology to finance and natural sciences}},
  edited by \bibinfo{editor}{\bibfnamefont{G.}~\bibnamefont{Caldarelli}}
  \bibnamefont{and}
  \bibinfo{editor}{\bibfnamefont{A.}~\bibnamefont{Vespignani}}
  (\bibinfo{publisher}{World Scientific}, \bibinfo{address}{Singapore},
  \bibinfo{year}{2007}), pp. \bibinfo{pages}{35--66}.

\bibitem[{\citenamefont{Hughes}(1995)}]{hughes}
\bibinfo{author}{\bibfnamefont{B.}~\bibnamefont{Hughes}},
  \emph{\bibinfo{title}{Random walks and random environments}}
  (\bibinfo{publisher}{Clarendon Press}, \bibinfo{address}{Oxford (UK)},
  \bibinfo{year}{1995}).

\bibitem[{\citenamefont{Lov{\'a}sz}(1996)}]{lovasz}
\bibinfo{author}{\bibfnamefont{L.}~\bibnamefont{Lov{\'a}sz}}, in
  \emph{\bibinfo{booktitle}{Combinatorics, Paul Erd{\"o}s is Eighty}}
  (\bibinfo{publisher}{J{\'a}nos Bolyai Mathematical Society, Budapest},
  \bibinfo{year}{1996}), p. \bibinfo{pages}{353}.

\bibitem[{\citenamefont{Dorogovtsev et~al.}(2000)\citenamefont{Dorogovtsev,
  Mendes, and Samukhin}}]{mendes99}
\bibinfo{author}{\bibfnamefont{S.~N.} \bibnamefont{Dorogovtsev}},
  \bibinfo{author}{\bibfnamefont{J.~F.~F.} \bibnamefont{Mendes}},
  \bibnamefont{and} \bibinfo{author}{\bibfnamefont{A.~N.}
  \bibnamefont{Samukhin}}, \bibinfo{journal}{Phys. Rev. Lett.}
  \textbf{\bibinfo{volume}{85}}, \bibinfo{pages}{4633} (\bibinfo{year}{2000}).

\bibitem[{\citenamefont{Pastor-Satorras
  et~al.}(2001)\citenamefont{Pastor-Satorras, V{\'a}zquez, and
  Vespignani}}]{alexei}
\bibinfo{author}{\bibfnamefont{R.}~\bibnamefont{Pastor-Satorras}},
  \bibinfo{author}{\bibfnamefont{A.}~\bibnamefont{V{\'a}zquez}},
  \bibnamefont{and}
  \bibinfo{author}{\bibfnamefont{A.}~\bibnamefont{Vespignani}},
  \bibinfo{journal}{Phys. Rev. Lett.} \textbf{\bibinfo{volume}{87}},
  \bibinfo{pages}{258701} (\bibinfo{year}{2001}).

\bibitem[{\citenamefont{Barrat and Pastor-Satorras}(2005)}]{barrat2005rea}
\bibinfo{author}{\bibfnamefont{A.}~\bibnamefont{Barrat}} \bibnamefont{and}
  \bibinfo{author}{\bibfnamefont{R.}~\bibnamefont{Pastor-Satorras}},
  \bibinfo{journal}{Phys. Rev. E} \textbf{\bibinfo{volume}{71}},
  \bibinfo{pages}{36127} (\bibinfo{year}{2005}).

\bibitem[{\citenamefont{Catanzaro et~al.}(2005)\citenamefont{Catanzaro,
  Bogu{\~n}\'a, and Pastor-Satorras}}]{catanzaro_ucm}
\bibinfo{author}{\bibfnamefont{M.}~\bibnamefont{Catanzaro}},
  \bibinfo{author}{\bibfnamefont{M.}~\bibnamefont{Bogu{\~n}\'a}},
  \bibnamefont{and}
  \bibinfo{author}{\bibfnamefont{R.}~\bibnamefont{Pastor-Satorras}},
  \bibinfo{journal}{Phys. Rev. E} \textbf{\bibinfo{volume}{71}},
  \bibinfo{eid}{027103} (\bibinfo{year}{2005}).

\bibitem[{\citenamefont{Bogu{\~n}{\'a}
  et~al.}(2004)\citenamefont{Bogu{\~n}{\'a}, Pastor-Satorras, and
  Vespignani}}]{mariancutofss}
\bibinfo{author}{\bibfnamefont{M.}~\bibnamefont{Bogu{\~n}{\'a}}},
  \bibinfo{author}{\bibfnamefont{R.}~\bibnamefont{Pastor-Satorras}},
  \bibnamefont{and}
  \bibinfo{author}{\bibfnamefont{A.}~\bibnamefont{Vespignani}},
  \bibinfo{journal}{Euro. Phys. J. B} \textbf{\bibinfo{volume}{38}},
  \bibinfo{pages}{205} (\bibinfo{year}{2004}).

\bibitem[{\citenamefont{Stauffer and Sahimi}(2005)}]{stauffer_annealed2005}
\bibinfo{author}{\bibfnamefont{D.}~\bibnamefont{Stauffer}} \bibnamefont{and}
  \bibinfo{author}{\bibfnamefont{M.}~\bibnamefont{Sahimi}},
  \bibinfo{journal}{Phys. Rev. E} \textbf{\bibinfo{volume}{72}},
  \bibinfo{pages}{46128} (\bibinfo{year}{2005}).

\bibitem[{\citenamefont{Almaas et~al.}(2003)\citenamefont{Almaas, Kulkarni, and
  Stroud}}]{almaas03:_scaling}
\bibinfo{author}{\bibfnamefont{E.}~\bibnamefont{Almaas}},
  \bibinfo{author}{\bibfnamefont{R.~V.} \bibnamefont{Kulkarni}},
  \bibnamefont{and} \bibinfo{author}{\bibnamefont{Stroud}},
  \bibinfo{journal}{Phys. Rev. E} \textbf{\bibinfo{volume}{68}},
  \bibinfo{pages}{056105} (\bibinfo{year}{2003}).

\bibitem[{\citenamefont{Hughes and Sahimi}(1982)}]{hughes82}
\bibinfo{author}{\bibfnamefont{B.~D.} \bibnamefont{Hughes}} \bibnamefont{and}
  \bibinfo{author}{\bibfnamefont{M.}~\bibnamefont{Sahimi}},
  \bibinfo{journal}{J. Stat. Mech.} \textbf{\bibinfo{volume}{29}},
  \bibinfo{pages}{781} (\bibinfo{year}{1982}).

\bibitem[{\citenamefont{Noh and Rieger}(2004)}]{nohrieger}
\bibinfo{author}{\bibfnamefont{J.}~\bibnamefont{Noh}} \bibnamefont{and}
  \bibinfo{author}{\bibfnamefont{H.}~\bibnamefont{Rieger}},
  \bibinfo{journal}{Phys. Rev. Lett.} \textbf{\bibinfo{volume}{92}},
  \bibinfo{pages}{118701} (\bibinfo{year}{2004}).

\bibitem[{\citenamefont{Abramowitz and Stegun}(1972)}]{abramovitz}
\bibinfo{author}{\bibfnamefont{M.}~\bibnamefont{Abramowitz}} \bibnamefont{and}
  \bibinfo{author}{\bibfnamefont{I.~A.} \bibnamefont{Stegun}},
  \emph{\bibinfo{title}{Handbook of mathematical functions.}}
  (\bibinfo{publisher}{Dover}, \bibinfo{address}{New York},
  \bibinfo{year}{1972}).

\bibitem[{\citenamefont{Gallos}(2004)}]{gallos04:_random_walk}
\bibinfo{author}{\bibfnamefont{L.~K.} \bibnamefont{Gallos}},
  \bibinfo{journal}{Phys. Rev. E} \textbf{\bibinfo{volume}{70}},
  \bibinfo{pages}{046116} (\bibinfo{year}{2004}).

\bibitem[{\citenamefont{{J. A. Holyst et al.}}(2005)}]{holyst05:_univer_scalin}
\bibinfo{author}{\bibnamefont{{J. A. Holyst et al.}}}, \bibinfo{journal}{Phys.
  Rev. E} \textbf{\bibinfo{volume}{72}}, \bibinfo{pages}{026108}
  (\bibinfo{year}{2005}).

\bibitem[{\citenamefont{Dorogovtsev et~al.}(2006)\citenamefont{Dorogovtsev,
  Mendes, and Oliveira}}]{dorogovtsev06:_degree_depen}
\bibinfo{author}{\bibfnamefont{S.~N.} \bibnamefont{Dorogovtsev}},
  \bibinfo{author}{\bibfnamefont{J.}~\bibnamefont{Mendes}}, \bibnamefont{and}
  \bibinfo{author}{\bibfnamefont{J.}~\bibnamefont{Oliveira}},
  \bibinfo{journal}{Phys. Rev. E} \textbf{\bibinfo{volume}{73}},
  \bibinfo{pages}{056122} (\bibinfo{year}{2006}).

\bibitem[{\citenamefont{Samukhin et~al.}(2008)\citenamefont{Samukhin,
  Dorogovtsev, and Mendes}}]{samukhin08:_laplac}
\bibinfo{author}{\bibfnamefont{A.~N.} \bibnamefont{Samukhin}},
  \bibinfo{author}{\bibfnamefont{S.~N.} \bibnamefont{Dorogovtsev}},
  \bibnamefont{and} \bibinfo{author}{\bibfnamefont{J.~F.~F.}
  \bibnamefont{Mendes}}, \bibinfo{journal}{Phys. Rev. E}
  \textbf{\bibinfo{volume}{77}}, \bibinfo{pages}{036115}
  (\bibinfo{year}{2008}).

\bibitem[{\citenamefont{Argyrakis and Kopelman}(2000)}]{argyrakis2000rwa}
\bibinfo{author}{\bibfnamefont{P.}~\bibnamefont{Argyrakis}} \bibnamefont{and}
  \bibinfo{author}{\bibfnamefont{R.}~\bibnamefont{Kopelman}},
  \bibinfo{journal}{Chemical Physics} \textbf{\bibinfo{volume}{261}},
  \bibinfo{pages}{391} (\bibinfo{year}{2000}).

\bibitem[{\citenamefont{Katsoulis et~al.}(2002)\citenamefont{Katsoulis,
  Argyrakis, Pimenov, and Vitukhnovsky}}]{katsoulis2002dat}
\bibinfo{author}{\bibfnamefont{D.}~\bibnamefont{Katsoulis}},
  \bibinfo{author}{\bibfnamefont{P.}~\bibnamefont{Argyrakis}},
  \bibinfo{author}{\bibfnamefont{A.}~\bibnamefont{Pimenov}}, \bibnamefont{and}
  \bibinfo{author}{\bibfnamefont{A.}~\bibnamefont{Vitukhnovsky}},
  \bibinfo{journal}{Chemical Physics} \textbf{\bibinfo{volume}{275}},
  \bibinfo{pages}{261} (\bibinfo{year}{2002}).

\bibitem[{\citenamefont{Redner}(2001)}]{redner_fpp}
\bibinfo{author}{\bibfnamefont{S.}~\bibnamefont{Redner}},
  \emph{\bibinfo{title}{A Guide to First-Passage Processes}}
  (\bibinfo{publisher}{Cambridge University Press}, \bibinfo{address}{Cambridge
  (UK)}, \bibinfo{year}{2001}).

\bibitem[{\citenamefont{Newman}(2003)}]{newmanrev}
\bibinfo{author}{\bibfnamefont{M.~E.~J.} \bibnamefont{Newman}}, in
  \emph{\bibinfo{booktitle}{Handbook of Graphs and Networks: From the Genome to
  the {I}nternet}}, edited by
  \bibinfo{editor}{\bibfnamefont{S.}~\bibnamefont{Bornholdt}} \bibnamefont{and}
  \bibinfo{editor}{\bibfnamefont{H.~G.} \bibnamefont{Schuster}}
  (\bibinfo{publisher}{Wiley-VCH}, \bibinfo{address}{Berlin},
  \bibinfo{year}{2003}), pp. \bibinfo{pages}{35--68}.

\bibitem[{\citenamefont{Baronchelli and Loreto}(2006)}]{baronchelli_rw2006}
\bibinfo{author}{\bibfnamefont{A.}~\bibnamefont{Baronchelli}} \bibnamefont{and}
  \bibinfo{author}{\bibfnamefont{V.}~\bibnamefont{Loreto}},
  \bibinfo{journal}{Phys. Rev. E} \textbf{\bibinfo{volume}{73}},
  \bibinfo{pages}{026103} (\bibinfo{year}{2006}).

\bibitem[{\citenamefont{V{\'a}zquez et~al.}(2003)\citenamefont{V{\'a}zquez,
  Bogu{\~n}\'a, Moreno, Pastor-Satorras, and Vespignani}}]{structurednets}
\bibinfo{author}{\bibfnamefont{A.}~\bibnamefont{V{\'a}zquez}},
  \bibinfo{author}{\bibfnamefont{M.}~\bibnamefont{Bogu{\~n}\'a}},
  \bibinfo{author}{\bibfnamefont{Y.}~\bibnamefont{Moreno}},
  \bibinfo{author}{\bibfnamefont{R.}~\bibnamefont{Pastor-Satorras}},
  \bibnamefont{and}
  \bibinfo{author}{\bibfnamefont{A.}~\bibnamefont{Vespignani}},
  \bibinfo{journal}{Phys. Rev. E} \textbf{\bibinfo{volume}{67}},
  \bibinfo{pages}{046111} (\bibinfo{year}{2003}).

\bibitem[{\citenamefont{Condamin et~al.}(2007)\citenamefont{Condamin,
  B\'enichou, Tejedor, Voituriez, and Klafter}}]{condamin_nature}
\bibinfo{author}{\bibfnamefont{S.}~\bibnamefont{Condamin}},
  \bibinfo{author}{\bibfnamefont{O.}~\bibnamefont{B\'enichou}},
  \bibinfo{author}{\bibfnamefont{V.}~\bibnamefont{Tejedor}},
  \bibinfo{author}{\bibfnamefont{R.}~\bibnamefont{Voituriez}},
  \bibnamefont{and} \bibinfo{author}{\bibfnamefont{J.}~\bibnamefont{Klafter}},
  \bibinfo{journal}{Nature} \textbf{\bibinfo{volume}{4501}},
  \bibinfo{pages}{77} (\bibinfo{year}{2007}).

\bibitem[{\citenamefont{Song et~al.}(2006)\citenamefont{Song, Havlin, and
  Makse}}]{song2006ofg}
\bibinfo{author}{\bibfnamefont{C.}~\bibnamefont{Song}},
  \bibinfo{author}{\bibfnamefont{S.}~\bibnamefont{Havlin}}, \bibnamefont{and}
  \bibinfo{author}{\bibfnamefont{H.}~\bibnamefont{Makse}},
  \bibinfo{journal}{Nature Physics} \textbf{\bibinfo{volume}{2}},
  \bibinfo{pages}{275} (\bibinfo{year}{2006}).

\bibitem[{\citenamefont{{Villas Boas} et~al.}(2007)\citenamefont{{Villas Boas},
  Rodrigues, Travieso, and Costa}}]{villasboas2007btc}
\bibinfo{author}{\bibfnamefont{P.}~\bibnamefont{{Villas Boas}}},
  \bibinfo{author}{\bibfnamefont{F.~A.} \bibnamefont{Rodrigues}},
  \bibinfo{author}{\bibfnamefont{G.}~\bibnamefont{Travieso}}, \bibnamefont{and}
  \bibinfo{author}{\bibfnamefont{L.}~\bibnamefont{Costa}},
  \emph{\bibinfo{title}{Border trees of complex networks}}
  (\bibinfo{year}{2007}), \bibinfo{note}{eprint arXiv:0706.3403v1}.

\bibitem[{\citenamefont{Watts and Strogatz}(1998)}]{watts_strogatz}
\bibinfo{author}{\bibfnamefont{D.~J.} \bibnamefont{Watts}} \bibnamefont{and}
  \bibinfo{author}{\bibfnamefont{S.~H.} \bibnamefont{Strogatz}},
  \bibinfo{journal}{Nature} \textbf{\bibinfo{volume}{393}},
  \bibinfo{pages}{440} (\bibinfo{year}{1998}).

\end{thebibliography}
\end{document}